\documentclass[fleqn,usenatbib]{mnras}

\usepackage{bm}
\usepackage{multirow}

\usepackage{newtxtext,newtxmath}
\usepackage[T1]{fontenc}

\DeclareRobustCommand{\VAN}[3]{#2}
\let\VANthebibliography\thebibliography
\def\thebibliography{\DeclareRobustCommand{\VAN}[3]{##3}\VANthebibliography}

\usepackage{graphicx}	
\usepackage{amsmath}	

\usepackage{amssymb}	
\usepackage{caption}
\usepackage{subcaption}

\graphicspath{{./}{images/}}

\title[Data-driven scale identification in oscillatory dynamos]{Data-driven scale identification in oscillatory dynamos}

\author[A. Guseva]{
Anna Guseva\thanks{E-mail: anna.guseva@obspm.fr}
\\
Department of Applied Mathematics,  University of Leeds, Leeds LS2 9JT, UK \\
LERMA, Observatoire de Paris, Paris, France
}

\date{Accepted XXX. Received YYY; in original form ZZZ}
\pubyear{2023}

\begin{document}
\label{firstpage}
\pagerange{\pageref{firstpage}--\pageref{lastpage}}
\maketitle


\begin{abstract}
    Parker's mean-field model includes two processes generating  large-scale oscillatory dynamo waves: stretching of magnetic field lines by small-scale helical flows, and by differential rotation. In this work, we investigate the capacity of data-driven modal analysis, Dynamic Mode Decomposition, to identify coherent magnetic field structures of this model. In its canonical form, the only existing field scale corresponds to the dynamo instability.  To take into account multi-scale nature of the dynamo, the model was augmented with coherent in time flow field, forcing small-scale magnetic field with a faster temporal evolution. Two clusters of  DMD modes were obtained: the ``slow" cluster, located near the dynamo wave frequency and associated with its nonlinear self-interaction, and the ``fast" cluster, centered around the forcing frequency and resulting from the interaction between the wave and the flow. Compared to other widely used methods of data analysis, such as Fourier transform, DMD provides a natural spatiotemporal basis for the dynamo, related to its nonlinear dynamics. We assess how the parameters of the DMD model, rank and delay, influence its accuracy, and finally discuss the limitations of this approach when applied to randomly forced, more complex dynamo flows.

\end{abstract}


\section{Introduction}\label{sec:intro}
Large-scale magnetic fields play a key role in formation and evolution of planets, stars and accretion discs. In the Sun and some low-mass stars, these fields evolve periodically in time, resulting in stellar magnetic cycles \citep{munoz2019visualization,saikia2016solar}.  Mean-field electrodynamics explains sustained large-scale astrophysical magnetic fields through systematic stretching and twisting of magnetic field lines by turbulent flows \citep{parker1955hydromagnetic,moffatt1978magnetic,krause1980mean}. This process is efficient when 
 fluctuating flow $u'$ and fluctuating field $b'$ correlate well, so that the mean electromotive force, or emf $\varepsilon = \langle u' \times b' \rangle$,  is non-zero. In mean-field theory, the emf is assumed as a truncated expansion in the large-scale magnetic field \citep{charbonneau2020dynamo}. In isotropic turbulent flows non-zero kinetic helicity, the first component of this expansion is $\varepsilon_i = \alpha \delta_{ij} B_j$. 
 In a nutshell, the dynamo can be described with the one-dimensional $\alpha-\Omega$ model, 
\begin{eqnarray}\label{eq:richproct}
A_t & = & \alpha B + \eta(A_{xx} - l^2 A), \label{eq:stand_A} \\
B_t & = & \Omega' A_x  - B^3 + \eta (B_{xx} - l^2 B), \label{eq:stand_B} 
\end{eqnarray}
where $\eta$ is magnetic diffusivity \citep{proctor2007effects,richardson2010effects}. The $x$-coordinate represents latitude or vertical direction in the flow; $l$ is an inverse length scale of the field in the other spatial directions. $\Omega$-effect corresponds to  stretching of  poloidal magnetic field with potential $A$ into toroidal field $B$ by large-scale differential rotation $\Omega'$; $\alpha$-effect parametrizes generation of $A$ from $B$ by helical turbulent motions.  Depending on the form of $\alpha$, equations (\ref{eq:richproct}-\ref{eq:stand_B}) admit steady or oscillatory linearly unstable solutions of  dipolar or quadrupolar parity \citep{richardson2010effects}. The oscillatory solutions, travelling latitudinally and resembling solar magnetic activity, are also known as \textit{dynamo waves}. Mean-field model (\ref{eq:richproct}-\ref{eq:stand_B}) is kinematic, i.e. magnetic field grows linearly from a prescribed flow parametrized by $\Omega'$ and $\alpha$. It can be extended to include shear in the solar tachocline \citep{charbonneau1997solar}, flux transport by meridional flows \citep{babcock1961topology,leighton1969magneto,charbonneau2014solar,cameron2018observing}, and dynamical feedback of the field on the flow through Lorentz force \citep{tobias1997solar,bushby2003strong}. Large-scale dynamo cycles observed in direct numerical simulations (DNS) of convective turbulence are often consistent with the mean-field theory \citep{schrinner2011oscillatory,racine2011mode,kapyla2013effects}.

Symmetry-preserving nonlinear term  $B^3$ was introduced into (\ref{eq:stand_B}) to model saturation of the dynamo instability to a steady state. This can happen due to expulsion of magnetic flux tubes from active dynamo regions by magnetic buoyancy \citep{parker1979cosmical}, the feedback from the Lorentz force slowing down differential rotation \citep{gilman1983dynamically} or reducing stretching properties of the flow ($\alpha$-effect) when magnetic energy becomes comparable to kinetic energy of the flow \citep[e.g.,][]{jones2010solar}. In the astrophysically relevant limit of weak magnetic diffusion, generation of magnetic field at small scales is favoured over large ones \citep[e.g.,][]{brandenburg2005astrophysical}, implying very low saturation levels for large-scale magnetic fields \citep{vainshtein1992nonlinear,cattaneo1996nonlinear}. However, numerical simulations of  \citet{tobias2013shear} showed that strong shear, developing in many astrophysical flows, promotes large-scale  dynamo waves by suppressing fluctuations at small scales. Shear may also facilitate temporally coherent magnetic patterns in self-sustained magnetorotational turbulence without helical or convective forcing \citep{nauman2016sustained}. 
Thus, dynamos organize at large scales through interactions between turbulence, shear, small-scale and  large-scale fields (see the reviews by \citet{brandenburg2018advances}, \citet{rincon2019dynamo}, \citet{tobias2021turbulent} and references therein). To understand these nonlinear interactions, it is necessary to  unambiguously identify coherent dynamo structures in multiscale dynamo models like state-of-art DNS (with about $10^6$ degrees of freedom).

Spatiotemporal coherence of dynamo waves implies that separated in space fluid parcels evolve synchronously while  waves propagate through the flow, and there is a particular spatial structure of magnetic and velocity fields associated with this wave. Some insight into this structure can be gained from Fourier transform or wavelet analysis; however, results  are inevitably biased towards \textit{a priori} choice of spatial and temporal basis inherent in these methods. Another way is to analyse eigenmodes of the dynamo, obtained by linearizing its equations about some $n$-dimensional basic state. However, this linearization is valid only in the vicinity of parameters where it was performed and does not necessarily capture nonlinear behaviour of the system away from this region.  A more general approach is to obtain these structures directly from  dynamo data using the methods of \textit{data-driven modal decomposition}. Frequently applied for analysis of hydrodynamic turbulence \citep{taira2020modal}, they have not been widely employed in astrophysics; yet they were found useful for torsional magnetic wave detection \citep{hori2023jupiter}. In essence, modal decomposition is a factorization of  the matrix of flow observables into matrices representing their spatial structure, temporal dynamics, and amplitudes indicating relative importance of every spatial mode.  Proper Orthogonal Decomposition (POD) produces modes that give an optimal representation of the flow in terms of energy compared to any other basis of the same dimension \citep{lumley2007stochastic,sirovich1987turbulence,holmes2012turbulence}; it can mix different time and length scales in a single mode if their energy contribution is comparable. Dynamic Mode Decomposition (DMD) seeks for the closest approximation of the flow data in terms of a linear system,  and is efficient for analysing periodic and quasi-periodic dynamics \citep{schmid2010dynamic,bagheri2010analysis,schmid2022dynamic}.  POD and DMD are entirely data-driven and therefore can be applied independently of boundary conditions and flow complexity.  Moreover, compared to empirical temporal analysis such as Hilbert-Huang transform \citep{huang1998empirical}, these methods also provide physically interpretable spatial structures. Unlike Fourier and wavelet transforms, no \textit{a priori} spatiotemporal basis is assumed; both spatial shape and temporal behaviour of the modes result from the data analysis. Assumptions about linearity and stationarity of the signals are relaxed for Dynamic Mode Decomposition, as compared to Fourier transform. As a drawback, DMD requires two empirical, user-defined parameters, rank and delay, and we will discuss below the influence of these parameters on decomposition accuracy. POD and DMD result in large modal bases for very chaotic signals; this is not the case for dynamo mean-field models analysed here.

In this paper, we will test the applicability of these methods for detection of scales and their interaction on two dynamo problems: Parker's $\alpha-\Omega$ model (\ref{eq:stand_A}-\ref{eq:stand_B}) and an extended dynamo model featuring both large and small scales. This paper is structured as follows: in section~\ref{sec:methods} we describe the data-driven approach, POD and DMD methods; in section~\ref{sec:RiPr_benchmark} we apply this methods to analysis of Parker's dynamo waves; in section~\ref{sec:RiPr_aug} we construct a more complex dynamo model with scale separation, analyse its dynamics with DMD. In section~5 we assess the accuracy of the DMD model. Section~\ref{sec:discussion} concludes the paper. 

\section{Data-driven approach}\label{sec:methods}
\begin{figure}
\includegraphics[width=\columnwidth]{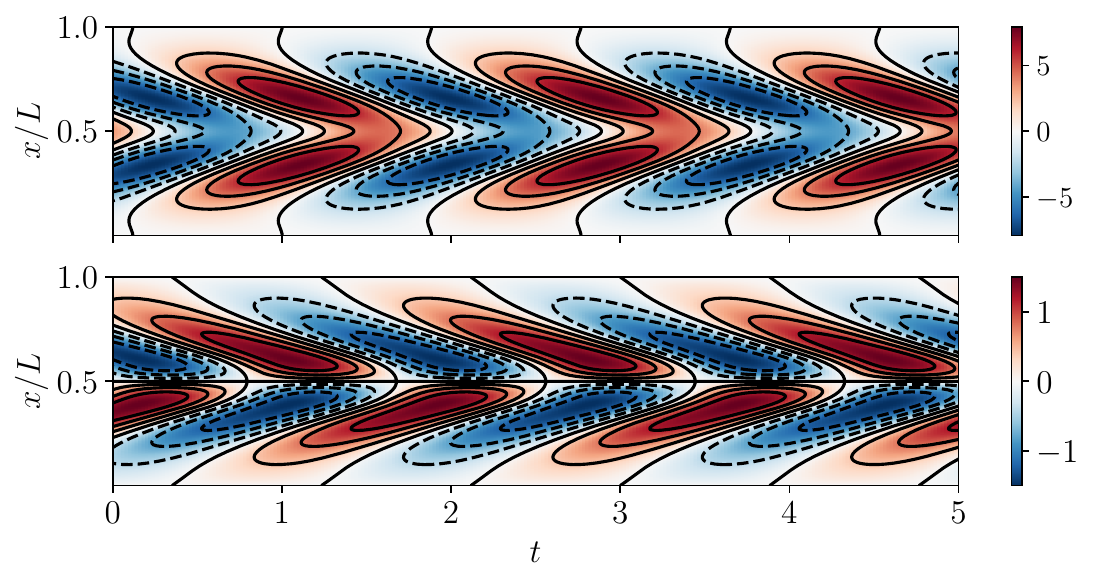}
\caption{Time evolution of dynamo waves in system~(\ref{eq:stand_A}-\ref{eq:stand_B}): magnetic potential $A$ (top), toroidal component of magnetic field $B$ (bottom). The data are collected until $t=100$; $D = 30$. \label{fig:richproct_waves}} 
\end{figure}
Consider the original nonlinear system (\ref{eq:stand_A}-\ref{eq:stand_B}) or an analogous one which evolves the magnetic potential $A$ and the magnetic field $B$ and depends on a set $\bm{\mu}$ of dimensionless functions and parameters,
\begin{equation}\label{eq:nonlsys}
\frac{d \bm{q}}{d t} = \bm{f}^{nonlinear} (\bm{q}, t, \bm{\mu}), \quad \bm{q}(t) = (A, B), \quad \bm{\mu} = \alpha, \Omega', \eta, l...
\end{equation}
The functional form of the $\alpha$-effect and the boundary conditions are
\begin{equation}\label{eq:richproct_alpha}
\alpha = - \alpha_0 \sin(2 \pi x/L), \quad A = B = 0 \text{ at }  x = 0,L=10.
\end{equation}
Dynamo number $D = \Omega' \alpha_0 / \eta^2 l^3$ is the dimensionless parameter of the system defining frequency of waves; if $\eta= l = \Omega' =1$, then $D=\alpha_0$. We solve system (\ref{eq:nonlsys}) numerically  in Python spectral solver Dedalus \citep{burns2020dedalus}, with Chebyshev discretization of $N=256$ points in  $x$ and Runge-Kutta time-stepping method.  The dynamo appears as a linear instability in the form of traveling waves for $D > 20$,  as shown in figure~\ref{fig:richproct_waves}. The initial conditions were of dipolar parity, with  $A$ symmetric and $B$ antisymmetric  in $x$. This way, the system is constrained to dipolar solutions; otherwise quadrupolar dynamos with antisymmetric $A$ and symmetric $B$ could also appear.

During the simulation, we collect instantaneous spatial profiles (snapshots) of the numerical solution of the system, $\bm{q}_k (x) \in R^N$, at time steps $k = 1,2, \dots, K$, with frequency sufficient to resolve the fastest flow dynamics, and assemble them into the data matrix $Q$,
\begin{equation}\label{eq:matrixQ}
  Q =[\bm{q}_1  \quad \bm{q}_2 \cdots \bm{q}_K],
\end{equation}
 for further analysis. Here, we performed POD and DMD analysis separately on the data for $A$ and $B$,  although treating $A$ and $B$ as a single system state  vector is also possible. 
 
\subsection{Proper Orthogonal Decomposition}\label{sec:POD}

Proper Orthogonal Decomposition (POD), also known as Principal Component Analysis, or Karhunen--Loeve expansion, is based on the idea that spatial flow correlations can be decomposed into orthogonal vectors ranked by their energy. Given system state $\mathbf{q}$, defined on the domain $\mathbf{x}$, POD modes are eigenfunctions of the integral equation
\begin{equation}\label{eq:PODint}
    \int \langle \mathbf{q (x)} \mathbf{q^* (x')} \rangle \phi(\mathbf{x'})\mathbf{dx'} = \sigma^2 \phi(\mathbf{x}),
\end{equation}
with autocorrelation function $\langle \mathbf{q(x)} \mathbf{q^*(x')} \rangle$ as a kernel, and $\langle \dots \rangle$ denoting temporal average.  The eigenvalues  $\sigma^2 \geq 0$ correspond to the average energy content in each mode $\phi$, and the first $r$ modes represent more of the system energy then any other spatial basis of order $r$.   A rigorous introduction to POD analysis for fluid flows can be found in \citet[pp. 86-128]{holmes2012turbulence}. For discretized data~(\ref{eq:matrixQ}), modes $\phi$ are eigenvectors of positive semi-definite correlation matrix $Q^* Q$. In practice, it is convenient to find $\phi$ by factorizing the data matrix $Q$
with Singular Value Decomposition (SVD),  
\begin{equation}\label{eq:POD}
    Q = \Phi \Sigma V^* \approx \Phi_r \Sigma_r V_r^*,
\end{equation}
which generalizes the eigenvalue decomposition to non-square matrices. The data set is thus represented as a product of the matrix $\Phi$ of the POD modes $\phi_i$, the matrix $\Sigma$ of their singular values $\sigma_i$, and their temporal evolution coefficients $V$. $Q$ is typically low-rank in fluid flows, so most of the singular values are nearly zero; the corresponding modes contain very little energy. A common practice is to retain a subset $r$ of modes with energy content larger than a threshold, and to discard the rest as noise.  To simplify our analysis, we interpolated all our data from non-equispaced Chebyshev points to an equispaced grid; otherwise, an energy weight matrix is required in~(\ref{eq:PODint},\ref{eq:POD}) \citep[see][]{schmidt2020guide}.

\subsection{Dynamic Mode Decomposition}\label{sec:DMD}
Unlike statistics-based POD, Dynamic Mode Decomposition (DMD) harnesses temporal dynamical behaviour of the flow.  It is closely related to Koopman operator theory, proposing that the temporal dynamics of nonlinear system (\ref{eq:nonlsys}) can be described with a linear operator acting on suitable flow observables \citep{koopman1931hamiltonian}. For an arbitrary nonlinear system, there is no generalized method of constructing an exhaustive set of such observables and the operator itself. DMD provides an approximation to eigenvectors and eigenvalues of Koopman operator from the data sequence, without calculating the operator explicitly  \citep{tu2014dynamic}. The reader is referred to the comprehensive review by \citet{schmid2022dynamic} for a rigorous derivation of the DMD algorithm and its links to the Koopman theory.

\subsubsection{Standard DMD}
In this work, we use the ``exact" definition of the DMD algorithm proposed by \citet{tu2014dynamic}. We seek to approximate nonlinear system (\ref{eq:nonlsys}), discretized in time,  with a linear operator $\mathcal{A}$ that maps the flow state at time $t_k$ to the flow state at time $t_{k+1}$,
\begin{equation}\label{eq:linsys_disc}
\bm{q}_{k+1} = \mathcal{A} \bm{q}_k, \quad \text{where} \quad \bm{q}_k = \sum_{i=1}^n \psi_i \lambda_i^k b_i(0),
\end{equation}
where $b_i(0)$ is the initial magnitude of the eigenvector $\psi_i$. Discrete eigenvalues $\lambda_i$ define whether  $\psi_i$ are decaying, neutral, or growing in time. We augment the data matrix (\ref{eq:matrixQ}) with one more snapshot of the system, and construct another matrix $Q'$, such that
  \begin{equation}\label{eq:linsys_data}
  Q' =[\bm{q}_2 \quad \bm{q}_2 \cdots \bm{q}_{K+1}],  \qquad Q' \approx \mathcal{A} Q.  
  \end{equation}
A least-squares approximation to $\mathcal{A}$ can be obtained by calculating the matrix product of the pseudoinverse of $Q$ with $Q'$. However, for low-rank matrices like $Q$, pseudoinverse is ill-conditioned and  leads to spurious results \citep{press1992numerical}. A workaround proposed by \citet{schmid2010dynamic} is to construct a lower, $r$-dimensional representation of matrix $Q$ by truncating its singular value decomposition (\ref{eq:POD}) at rank $r$.  After this operation, (\ref{eq:linsys_data}) is re-arranged as 
  \begin{equation}\label{eq:Ar}
  Q' \approx  \mathcal{A} \Phi_r \Sigma_r V_r^*, \qquad \Phi_r^* \mathcal{A} \Phi_r = \Phi_r^* Q' V_r \Sigma^{-1}_r = \mathcal{A}_r,
  \end{equation}
  where $\mathcal{A}_r$ is a projection of $\mathcal{A}$ on the POD modes $\Phi_r$. Its  eigenvalues and eigenvectors can be calculated from $\mathcal{A}_r \tilde{\psi} = \lambda \tilde{\psi}$.  The full-state eigenvectors in the physical space can be recovered  from $r$-dimensional $\tilde{\psi}$ through
  \begin{equation}\label{eq:DMD_modes}
 \psi = \frac{1}{\lambda} Q' V_r \Sigma^{-1}_r \tilde{\psi}, \qquad \omega = \ln(\lambda)/ \Delta t.
 \end{equation}
The second equation in (\ref{eq:DMD_modes}) relates discrete-time eigenvalues $\lambda$ to continuous-time eigenvalues $\omega$, both complex. In the following, we will refer to $\psi$ and $\omega$ as DMD modes and DMD eigenvalues. If the real part of the eigenvalue $\Re(\omega)<0$, the mode is dampened; if $\Re(\omega)>0$, the mode will be growing. In a steady state system, its significant modes are expected to be nearly neutral, $\Re(\omega)\approx 0$. The imaginary part $\Im(\omega)$ is the temporal frequency of the mode. 

As in section \ref{sec:POD}, threshold $r$ is user-defined. A common choice for systems with little noise is to retain first $r$ modes so that $\sum_r \sigma_r /\sum_{tot} \sigma = 0.99$, retaining 99\% information from the original data set. An alternative cut-off criterion is $\sum^2_r \sigma^2_r /\sum^2_{tot} \sigma^2 = 0.99$,  bearing in mind that $\sigma^2$, and not $\sigma$, represents the energy of POD modes in (\ref{eq:PODint}). The latter criterion means that 99\% of energy is retained and results in a sparser spatial basis. The relative error of the DMD model can be estimated as 
\begin{equation}\label{eq:lin_error}
   \epsilon =  ||Q - Q^{model}||_2/||Q||_2,
\end{equation}
where $||\cdot||$ is $L_2$-norm and $Q^{model}$ is the approximation to $Q$ calculated according to~(\ref{eq:linsys_disc}).

\subsection{High-order (Hankel) DMD}\label{sec:HankelDMD}
The robustness of DMD in ergodic dynamical systems can be improved by embedding state vectors $q_k$ in a higher dimension.  Using the method of delays  \citep{takens1981detecting}, we can construct Hankel matrix $Q_H$ from the original data~(\ref{eq:matrixQ}) as 
\begin{equation}\label{eq:HankelQ}
Q_H = 
\left[
\begin{array}{cccc}
    \bm{q}_1 & \bm{q}_2 & \cdots & \bm{q}_{K-d + 1} \\
    \bm{q}_2 & \bm{q}_3 &  \cdots  &  \bm{q}_{K-d + 2} \\
     \vdots &  \vdots & & \vdots  \\
    \bm{q}_{d} & \bm{q}_{d+1} &  \cdots & \bm{q}_{K} \\
\end{array} \right],
\end{equation}
where $d$ is the delay parameter, and augment (\ref{eq:linsys_data}) in the same way. This is done because our initial choice of system observables, $\bm{q}_k$, is not necessarily contained within the finite-dimensional invariant subspace of the Koopman operator. In Hankel matrix~(\ref{eq:HankelQ}), we increased the number of observables:  the matrix rank of $Q_H$ is higher then the rank of $Q$. \citet{arbabi2017ergodic} proved that in the limit $d\to \infty$ DMD eigenvalues and eigenmodes approximating the system 
  \begin{equation}\label{eq:linsys_data_Hankel}
Q'_H \approx \mathcal{A} Q_H.    
  \end{equation}
converge to Koopman modes and eigenvalues from the finite-dimensional invariant subspace of $\mathcal{A}$. Hankel DMD is also viewed as a higher-order approximation of~(\ref{eq:linsys_data}) \citep{le2017higher}.

Construction of (\ref{eq:HankelQ}) introduces another user-defined, delay parameter $d$, and its choice is not straightforward. \citet{takens1981detecting} stated that an attractor of dimension $m$ can be embedded into a space with $d = 2 m +1$ dimensions. 
 \citet{broomhead1986extracting}, considering physically relevant time scales of the system, suggested to choose $d$ so that the SVD spectrum of (\ref{eq:HankelQ}) converges as  the window length of delay $t_{k+d} - t_{k}$ is varied. In practice, $d$ is increased until the desired accuracy of approximation is achieved  in the sense of $L_2$-norm  \citep{fujii2019data}.

\section{Parker's dynamo waves}\label{sec:RiPr_benchmark}

As a benchmark problem, we take the system~(\ref{eq:stand_A}-\ref{eq:stand_B}) augmented with (\ref{eq:richproct_alpha}), as described in the beginning of section~\ref{sec:methods}. We analyse the steady-state data sets of toroidal magnetic field $B$ and poloidal magnetic potential $A$, depicted in figure~\ref{fig:richproct_waves}. 
\begin{figure*}
     \centering
     \begin{subfigure}[b]{0.45\textwidth}
         \centering
         \includegraphics[width=\textwidth]{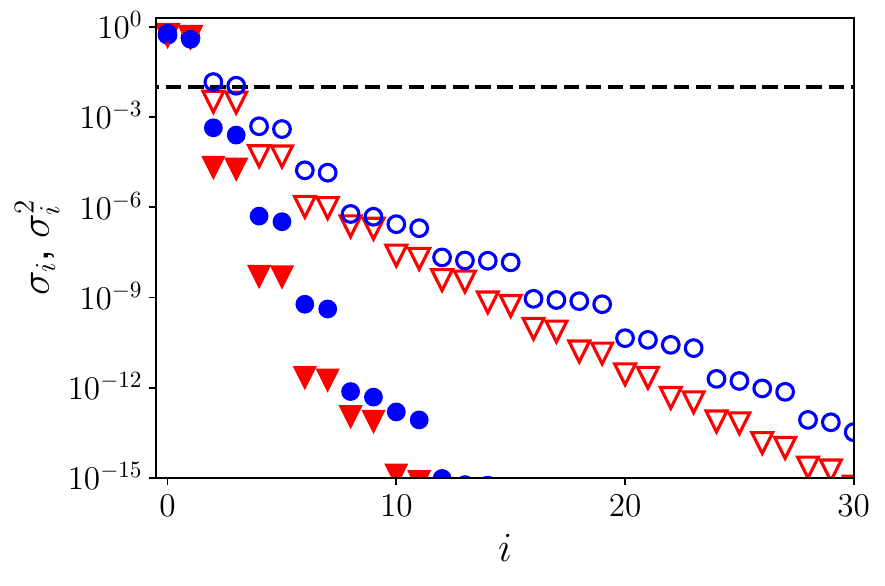}
         \caption{}
     \end{subfigure}
     \hfill
     \begin{subfigure}[b]{0.45\textwidth}
         \centering
         \includegraphics[width=\textwidth]{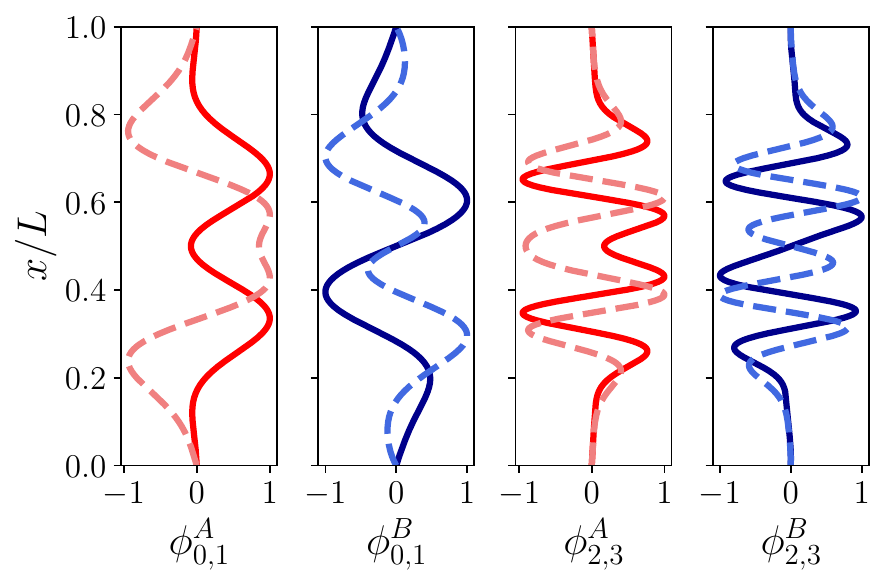}
         \caption{}
     \end{subfigure}
     \vfill 
          \begin{subfigure}[b]{0.45\textwidth}
         \centering
         \includegraphics[width=\textwidth]{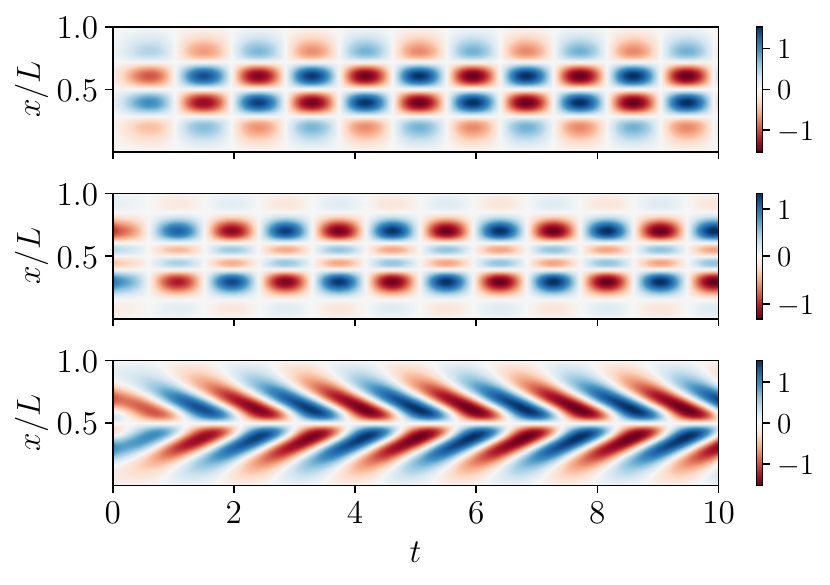}
         \caption{}
     \end{subfigure}
     \hfill
     \begin{subfigure}[b]{0.45\textwidth}
         \centering
         \includegraphics[width=\textwidth]{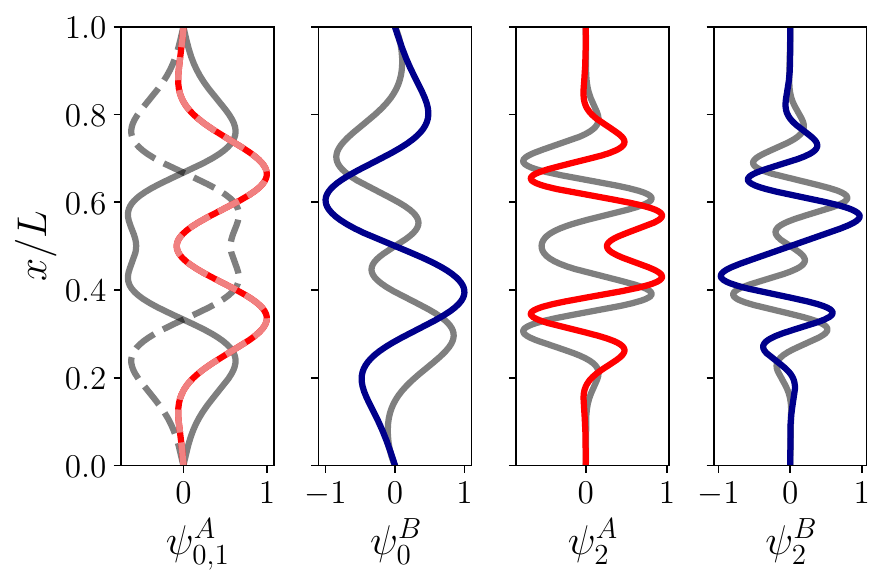}
         \caption{}
     \end{subfigure}
        \caption{(a) Singular value spectrum of the dynamo waves in figure~\ref{fig:richproct_waves}, system~(\ref{eq:stand_A}-\ref{eq:stand_B}). Empty symbols, $\sigma_i/\sum_i \sigma$, filled symbols, $\sigma^2_i/\sum_i \sigma^2$. Dashed line represent 99\% cut-off for both $\sigma$- and $\sigma^2$-criteria. (b) The first four POD modes of dynamo waves. Solid line, the first of the POD mode pairs in panel (a); dashed: the second mode in the pair.  (c)  Reduced model of the dynamo waves in figure~\ref{fig:richproct_waves} based on: top, $\phi^B_0$; middle, $\phi^B_1$, bottom: $\phi^B_0$ and $\phi^B_1$ together.
(d) DMD modes of the dynamo waves, calculated using the first four POD modes from panel (c). Solid lines: the first mode of the complex-conjugate modal pair. The real part of the mode is represented by a darker color and the imaginary part by a lighter (gray) color. Dashed, the same but for the second mode $\psi_1$ in the pair $\psi^*_1 = \psi_0$ (only for $\psi^A_{0,1}$). }
        \label{fig:RiPr_modes}
\end{figure*}


\subsection{POD decomposition}\label{sec:RiPr_POD}

Figure~\ref{fig:RiPr_modes}(a) shows the singular value spectrum of $A$ and $B$. The spectrum decays rapidly, indicating the low rank of the data and absence of noise. $B$ has more complex structure, with $4$ modes needed to describe 99\% of the data set according to $\sigma$-criterion, compared to $2$ modes for $A$. This difference can be attributed to the nonlinear term  $B^3$ in equation (\ref{eq:stand_B}), and therefore more complex dynamics. Potential $A$ is influenced by this nonlinearity only indirectly through $\alpha$-effect. According to $\sigma^2$-criterion, 99\% of energy in both $A$ and $B$ is captured by the first two modes. The  singular values of $A$ and $B$ come in nearly equal pairs, a common feature in fluid flows with  symmetries \citep{deane1991low}.  The dipolar symmetry implies that
\begin{equation}\label{eq:dip_symm}
    x \to -x, \quad A \to A, \quad B \to -B.
\end{equation}
 If the solution of (\ref{eq:nonlsys}) is invariant under the action of symmetry group $G$, and $\phi (x)$ is an eigenfunction of (\ref{eq:PODint}), then for every element $g\in G$ such as (\ref{eq:dip_symm}), $g \circ \phi$ is also an eigenfunction of (\ref{eq:PODint}) with the same eigenvalue  \citep[Proposition 3]{aubry1993preserving}.

The POD modes, corresponding to the singular value pairs, are shown in Figure~\ref{fig:RiPr_modes}(b). Following the fields parity in figure~\ref{fig:richproct_waves}, the $A$-modes $\phi^A_{0-3}$ are symmetric, and $B$-modes $\phi^B_{0-3}$ are antisymmetric with respect to $x$. The most energetic modes  $\phi^A_{0,1}$,  $\phi^B_{0,1}$ are large-scale and correspond to the length scale of the dynamo wave  observed in figure~\ref{fig:richproct_waves}. The spatial scale of the second, less energetic mode pair is approximately two times smaller, and its temporal dynamics is faster. Each mode in the pair $\phi^B_{0,1}$ (or $\phi^A_{0,1}$), weighted by its temporal coefficient (see equation~\ref{eq:inst_cI} in section~\ref{sec:lin_validity}), corresponds to a standing wave shifted in phase with respect to the other (figure~\ref{fig:RiPr_modes}c, upper and middle panel). They enhance each other at some spatial locations, and weaken at others, producing travelling wave dynamics observed in figure~\ref{fig:RiPr_modes} (c), bottom panel. The relative error between this 2-mode model and toroidal field $B$ in figure~\ref{fig:richproct_waves}, evaluated as $L_2$ norm, is only 3.7\%.

\subsection{DMD decomposition}\label{sec:DMD_parkers}
We calculate the DMD eigenmodes and eigenvalues of the dynamo waves in figure~\ref{fig:richproct_waves} using the algorithm in section~\ref{sec:DMD}. The previous section shows that truncation of rank $r=4$ is sufficient to reproduce the data  reliably. As they are real, the matrix $\mathcal{A}_r$ in~(\ref{eq:Ar}) is also real and its eigenvalues come in complex conjugate pairs with nearly zero real parts, as shown in table~\ref{tab:DMD_eigs_RiPr}.

The first frequency pair, $\Im(\omega_{0,1}) = \pm 3.558$, corresponds to the frequency of dynamo linear instability at $D=30$, and is the main frequency observed in figure~\ref{fig:richproct_waves}.
The second frequency is three times larger than the first one, $\Im(\omega_{2,3}) \approx 3 \Im(\omega_{0,1})$, and can be interpreted as an interaction of the first mode with itself through nonlinear damping $B^3$ in (\ref{eq:stand_B}), as the periodic temporal behaviour implies $\sin^3(\omega) \sim \sin(3\omega)$. Note the implication of choosing $\sigma$ or $\sigma^2$ cut-off: if only 2 POD modes are retained according to $\sigma^2$-criterion, the dynamics will be projected on the most energetic, slowly oscillating modes, and  the fast-oscillating modes  will not be recovered. The choice of cut-off $r$ is therefore a balance between desired accuracy and basis sparsity.

The first two panels in figure~\ref{fig:RiPr_modes}(d) show the ``slow" DMD modes of $A$ and $B$; one can observe that the $x$-profile of the real parts of modes is similar to one of the corresponding POD mode pairs in figure~\ref{fig:RiPr_modes}(b), while their imaginary part is resembles the other.  Like the eigenvalues, the DMD modes are also complex conjugates, $\psi^*_1 = \psi_0$, thus, their real parts are equal, and their imaginary parts have equal magnitude along $x$ but opposite signs. This allows to represent the travelling wave structures similar to the lower panel in figure~\ref{fig:RiPr_modes}(c)  with just one complex DMD mode $\psi_0$ and its temporal coefficient. Indeed, consider a DMD mode $\psi_0 (x) = \psi^{r}  + \psi^{i} \mathrm{i}$, and the corresponding projection of the data set on this mode, $c_0 (t) = c^{r}  + c^{i} \mathrm{i}$. One-mode approximation for the data matrix $Q$ would be
\begin{equation}\label{eq:DMD_1mode_approx}
    Q  \approx \psi^r c^r - \psi^i c^i +  (\psi^i c^r + \psi^r c^i) \mathrm{i}.  
\end{equation}
When the corresponding contribution from the conjugate mode $\psi_1$ will be added, the imaginary part of  (\ref{eq:DMD_1mode_approx}) will be cancelled, as the data matrix $Q$ is real;  the real part will be double of that of (\ref{eq:DMD_1mode_approx}). Thus, it is enough to track one  mode from the complex-conjugate pair in $A$ and $B$, and DMD provides a more compact vector basis for this system compared to POD. In the following, we will focus on the DMD approach. 

\begin{table}
    \centering
    \begin{tabular}{c|c|c}
      Mode & $\omega^A$ & $\omega^B$  \\ \hline
        $0$, $1$ & $-7.401 \cdot10^{-7} \pm  3.558 \mathrm{i}$  & $-9.525 \cdot10^{-6} \pm  3.558 \mathrm{i}$  \\
        $2$, $3$ &  $-1.063 \cdot10^{-4} \pm  10.673 \mathrm{i}$  & $-1.257 \cdot10^{-3} \pm  10.675 \mathrm{i}$
    \end{tabular}
    \caption{DMD eigenvalues of potential $A$ and toroidal magnetic field $B$, obtained based on $r=4$ truncation of the data. The modes are ranked by their growth rates (less decaying first). The eigenvalues were rounded up to third digit.}
    \label{tab:DMD_eigs_RiPr}
\end{table}

\section{Multi-scale dynamo model}\label{sec:RiPr_aug}

Parker's dynamo model (\ref{eq:stand_A}-\ref{eq:stand_B}) parametrises interaction of small-scale dynamo and turbulence with $\alpha$-effect. Thus, it has only one independent time and length scale, related to the dynamo linear instability at large scales. The second, faster and smaller length scale, identified by POD and DMD, is the result of magnetic field interacting with itself. It is also very weak, less than 1\% of the total magnetic energy, as can be seen in figure \ref{fig:RiPr_modes}(a). However, it is not the case for more realistic numerical models of dynamo where waves arise naturally from shear turbulence. In those simulations, the flow is driven by a combination of large-scale shear and a small-scale forcing, coherent or randomized, and  energy is distributed on a range of scales \citep{tobias2013shear, pongkitiwanichakul2016shear,nigro2017large}.  More advanced dynamo models allow to integrate large and small-scale dynamics simultaneously, for example, with a shell model describing 
 behaviour of velocity and magnetic fields  at small scales \citep{nigro2011study}. In line with such models, we augment system~(\ref{eq:stand_A}-\ref{eq:stand_B}) with extra terms, mimicking the multi-scale flow behaviour in the DNS, and study whether DMD can recognize mixed small and large scales of the system. The augmented one-dimensional kinematic dynamo equations are
 \begin{eqnarray}
 A_t & = & \varepsilon + \eta(A_{xx} - l^2 A), \label{eq:sys2_A}\\
 B_t & = & \Omega' A_x -  B^3 + \eta(B_{xx} - l^2 B), \label{eq:sys2_B} \\
  b_t & = & \mu u_x B  - \gamma b^3 + \eta(b_{xx} - l^2 b), \label{eq:sys2_b}\\
\varepsilon &= & \varepsilon_0 g(u b), \quad g(f) = \frac{1}{\sqrt{2\pi} \sigma} \exp(-\frac{x^2}{2 \sigma^2}),  \label{eq:sys2_emf} \\ 
    u& = & u_0 \sin(2 \pi k_u x / L \pm \omega_u t).\label{eq:sys2_u} 
\end{eqnarray}
Temporal and spatial evolution of the small-scale flow field  $u$ is prescribed by~(\ref{eq:sys2_u}).  Equation~(\ref{eq:sys2_b}) describes the evolution of small-scale fluctuations $b$ of toroidal magnetic field and is inspired by the full induction equation for magnetic fluctuations, see e.g. \citet[][p. 29]{tobias2021turbulent}. Since the latter is nonlinear in fluctuations due to the terms $\nabla \times (u' \times b' - \overline{u' \times b'})$, nonlinearity affects all scales of magnetic field, and dynamo should be able to saturate at both small and large scales in statistically steady flows. Exact mechanism of this saturation is still a subject of active research \citep{moffatt2019self}. For simplicity, we introduce the same form of nonlinearity  $\gamma b^3$ into~(\ref{eq:sys2_b}), which allows to control the saturation amplitude of $b$.  The dissipative term    in~(\ref{eq:sys2_b}) is proportional to $b_{xx}$; $ \mu u_x B$ models generation of $b$ through interaction of the flow and the large-scale field. Parameters $\gamma$ and $\mu$ are introduced to control the strength of nonlinearity and the induction-like term, respectively, in this simplified mathematical model of a multiscale dynamo. In a more realistic DNS simulation,  signals from $B$ and $b$ would be mixed in multiscale magnetic field, $B_s = B +b$. The interaction between $b$ and  $u$  contributes to fluctuating electromotive force, proportional to $u b $.  
Net emf $\varepsilon = \varepsilon_0 g( u b )$ on the scale of the dynamo waves in (\ref{eq:sys2_A}) is calculated by filtering product $ub$ with a Gaussian filter (\ref{eq:sys2_emf}) and ensuring $\varepsilon=0$ at the boundaries $x=0,L$. Alternatively, time averaging or envelope smoothing can be employed to model action of emf on large scales. 

\begin{figure*}
     \centering
     \begin{subfigure}[b]{0.45\textwidth}
         \centering
         \includegraphics[width=\textwidth]{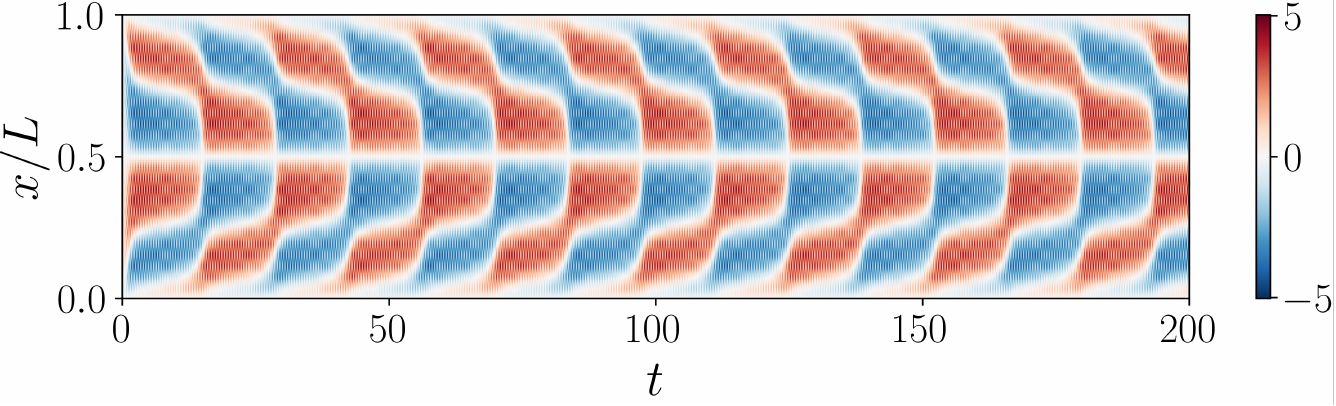}
         \caption{}
     \end{subfigure}
     \hfill
    \begin{subfigure}[b]{0.45\textwidth}
         \centering
         \includegraphics[width=\textwidth]{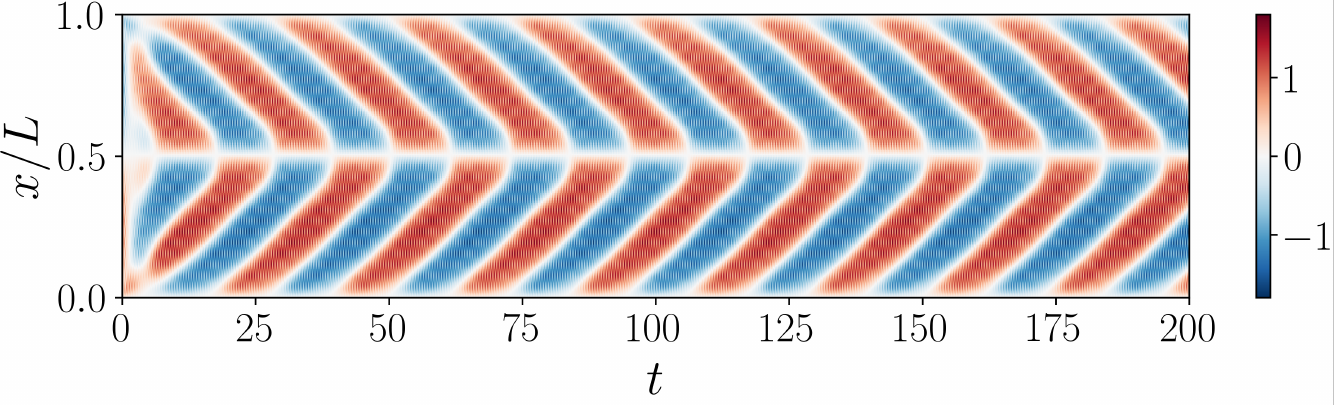}
         \caption{}
     \end{subfigure}

     \vfill 
    \begin{subfigure}[b]{0.45\textwidth}
         \centering
         \includegraphics[width=\textwidth]{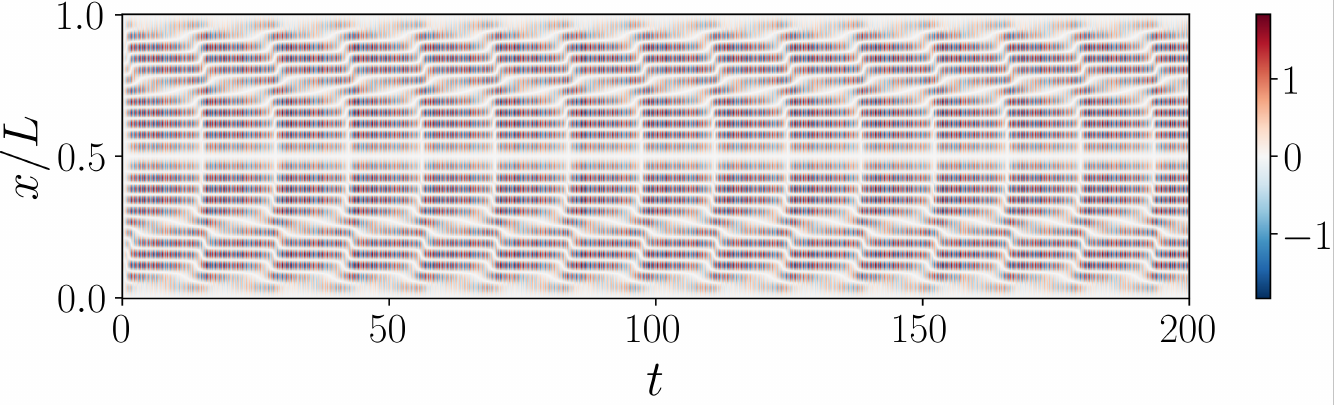}
         \caption{}
     \end{subfigure}
     \hfill
          \begin{subfigure}[b]{0.45\textwidth}
         \centering
         \includegraphics[width=\textwidth]{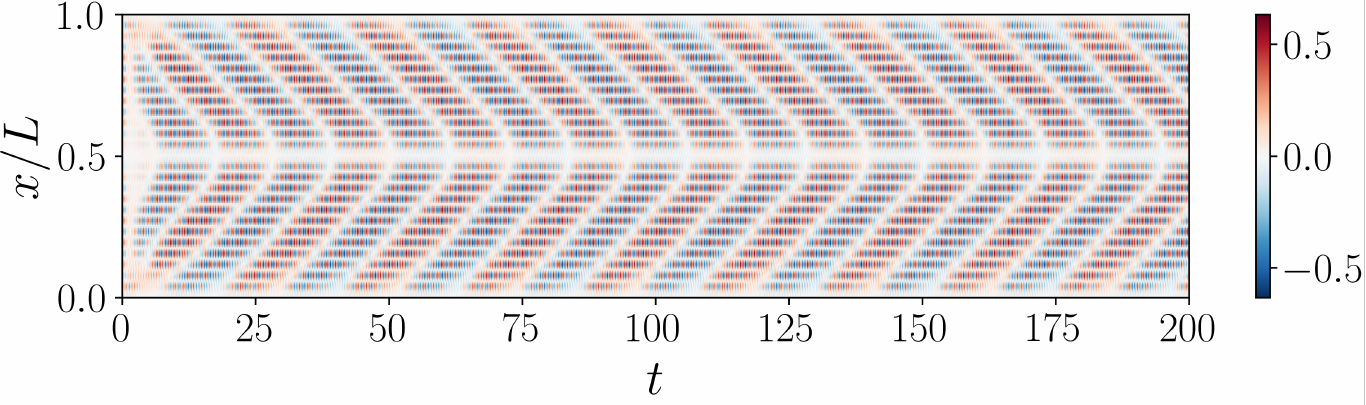}
         \caption{}
     \end{subfigure}
        \caption{The data obtained by integrating in time multi-scale dynamo equations (\ref{eq:sys2_A}-\ref{eq:sys2_b}). Without imposed $\alpha$-effect: multi-scale toroidal magnetic field $B_s = B+b$ (a) and  fluctuating toroidal magnetic field $b$ (b); flow  and emf given by~(\ref{eq:sys2_emf}-\ref{eq:sys2_u}). System with controlled $\alpha$-effect: $B_s$~(c),  $b$~(d);  flow and emf from (\ref{eq:RiPr_u}-\ref{eq:RiPr_aug_emf}). See text for description of the system parameters.}
        \label{fig:RiPr_kx13_dataset}
\end{figure*}

Dynamo waves were consistently observed when $\varepsilon_0 >800$ and the rest of parameters set to $\Omega' = \eta = l = 1$, $\mu=5$, $u_0 = 0.5$.  We set $\gamma=0$, allowing the dynamo to saturate only through large scales. Large value of $\sigma=30$  removes completely small-scale fluctuations from $\varepsilon$ and thus inhibits non-oscillatory dynamo solutions \citep{richardson2010effects}. Frequency $\omega_u$ and wave number $k_u$ of small-scale  $u$ were set to $\omega_u =13$, $k_u =13$ to ensure a faster temporal dynamics with respect to the dynamo wave and at least a decade of scale separation between the wave and the small-scale dynamo.  Figures~\ref{fig:RiPr_kx13_dataset}(a,c) show results of numerical integration of (\ref{eq:sys2_A}-\ref{eq:sys2_u}) with these parameters and $\varepsilon_0 = 1500$. 
The large-scale field $B$ is again dipolar, anti-symmetric with respect to the equator; it is the dominant feature of the dynamo waves observed in the total magnetic field $B_s$.  However, it is spatially and temporally modulated, allowing locally wider regions of positive and negative field direction. These differences appear because spatial distribution of net emf departs from the one in ``classic" dynamo waves. On the other hand, the fluctuating field $b$ in figure~\ref{fig:RiPr_kx13_dataset}(c) is predominantly small-scale, but yet with a systematic  footprint from the large-scale $B$, symmetric with respect to the $x/L=0.5$. Fast Fourier Transform (FFT) of multiscale magnetic field $B_s$ over the temporal domain, with  resulting spatial Fourier modes integrated in $x$, shows that the temporal dynamics in this system is complex and features numerous frequencies (figure~\ref{fig:RiPr_kx13_modes}a).   The most prominent peaks are clustered around the dominant dynamo ingredients: the ``slow" cluster corresponding to the dynamo wave and the ``fast" cluster  around the flow frequency $\omega_u$. Other degrees of freedom are excited in between these peaks, so the spectrum in the ``slow" cluster is relatively broad-band, which complicates identification of dynamo waves, small scales, and their interaction. It is thus of interest to construct a multiscale dynamo system featuring a well-defined $\alpha$-effect as in~\eqref{eq:richproct_alpha}.  

\subsection{Imposing $\alpha$-effect}\label{sec:aug_sin_alpha} 
For this, we add one more component to the flow velocity,
 \begin{equation}\label{eq:RiPr_u}
   u =  u_0 (\sin(\pi x/L) + \sin(2 \pi k_u x / L \pm \omega_u t)),
\end{equation}
so that magnetic fluctuations scale as $b\sim \cos(\pi x/L) B$. If  electromotive force is defined as
\begin{equation}\label{eq:RiPr_aug_emf}
 \varepsilon = - \varepsilon_0 g(u b) \sim \sin(2 \pi x/L) B,    
\end{equation}
then $\alpha$-effect is similar to~\eqref{eq:richproct_alpha}, yet dynamo waves appear from interaction of the flow and small-scale magnetic field.  Formally, the extra component in~(\ref{eq:RiPr_u}) can be interpreted as a large-scale one, although it is not constant in $x$ compared to mean differential rotation $\Omega'$ in~\eqref{eq:sys2_B}. Magnetic fluctuations will thus contain direct interaction of this component in~\eqref{eq:RiPr_u} and large-scale magnetic field in induction equation~(\ref{eq:sys2_b}), unlike in the mean-field theory. This inconsistency is resolved in the previous multiscale dynamo system (\ref{eq:sys2_A}-\ref{eq:sys2_u}) without imposed $\alpha$-effect, where large-scale shear interacts with $B$ indirectly through  $\Omega' A_x$ and $u_x B$.
Controlling $\alpha$-effect in this way allows to obtain a multiscale dynamo with few degrees of freedom, where large scales are similar to Parker's dynamo waves~(\ref{eq:stand_A}-\ref{eq:stand_B}).

We perform a numerical simulation of augmented model~(\ref{eq:sys2_A}-\ref{eq:sys2_b}), together with~(\ref{eq:RiPr_u}-\ref{eq:RiPr_aug_emf}), with dipolar initial conditions. Model parameters were set to $\varepsilon_0 = 30$, $\Omega' = l= \mu = 1$, $\eta = 0.1$,  $u_0=0.5$, $\sigma=10$, $\omega_u =13$, $k_u =13$, resulting in ``classic'' large-scale dynamo waves with a period $T \approx 20$, about ten times longer (figure~\ref{fig:RiPr_kx13_dataset}b). The small-scale nonlinearity was set non-zero, $\gamma = 10$, in order to reduce the magnitude of $b$ to values below those of large-scale $B$ (figure~\ref{fig:RiPr_kx13_dataset}d), as in dynamo without imposed $\alpha$-effect. 
The data were collected up to $T=200$, covering about  $10$ periods of wave evolution and thus resolving both slow and fast dynamics.  
 As previously, $b$ is mostly small-scale with a symmetric footprint of the large-scale field, arising through forcing of $b$ with the product of anti-symmetric $u_x$ and anti-symmetric $B$. Note the difference with multiscale dynamo system (\ref{eq:sys2_A}-\ref{eq:sys2_u}): the sign of the large-scale footprint in $b$ is periodic in time, following the sign of $u_x B \sim \cos(\pi x/L) B$. This is less apparent in figure~\ref{fig:RiPr_kx13_dataset}(c) because the product of $B$ and $u_x$ remains small-scale. The resulting multi-scale magnetic field, $B_s = B + b$ is thus of mixed parity with respect to $x$; however, since the  magnitude of $b$ is about $2$ times smaller than the one of $B$ and contains a significant energy contribution from small scales, this asymmetry very weak. FFT transform of $B_s$ in time shows much simpler temporal dynamics, with just a few dominant frequency peaks (figure~\ref{fig:RiPr_kx13_modes}b). In the following,  we will analyse this mixed signal $B_s$ with DMD, discarding transients up to $t=50$.

\subsection{DMD of the model with imposed $\alpha$-effect}\label{sec:RiPr_aug_DMD}

\begin{figure*}
     \centering
          \begin{subfigure}[b]{0.48\textwidth}
         \centering
         \includegraphics[width=\textwidth]{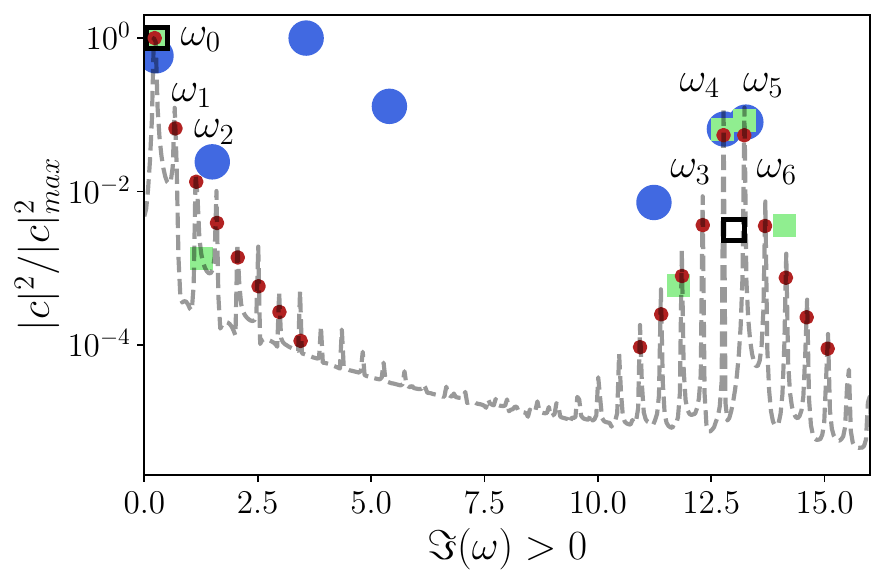}
         \caption{}
     \end{subfigure}
     \hfill 
          \begin{subfigure}[b]{0.48\textwidth}
         \centering
         \includegraphics[width=\textwidth]{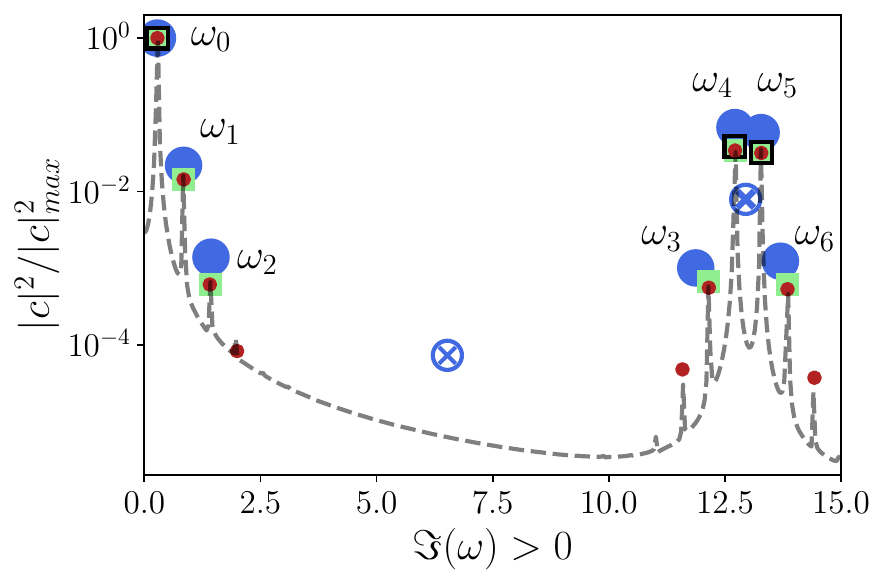}
         \caption{}
     \end{subfigure}

     \vfill 
     \begin{subfigure}[b]{0.48\textwidth}
         \centering
         \includegraphics[width=\textwidth]{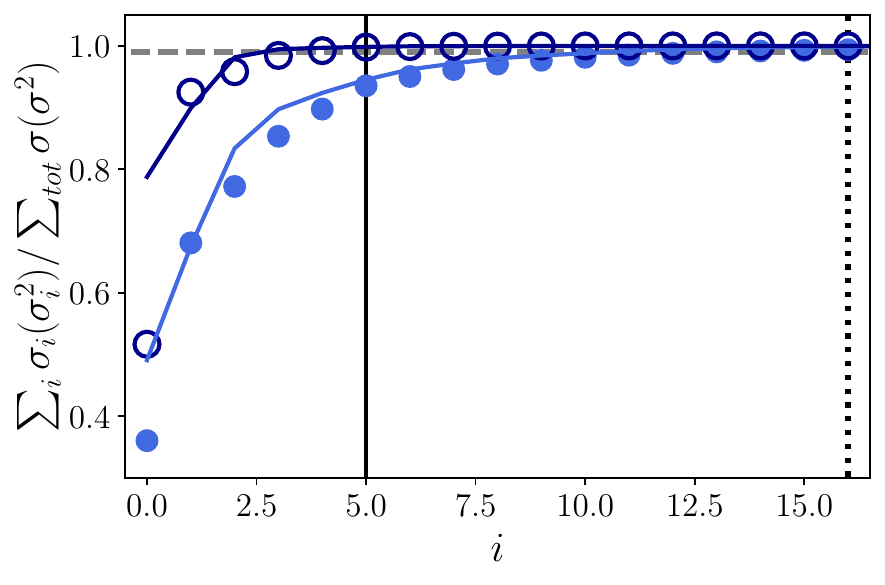}
         \caption{}
     \end{subfigure}
     \hfill 
          \begin{subfigure}[b]{0.48\textwidth}
         \centering
         \includegraphics[width=\textwidth]{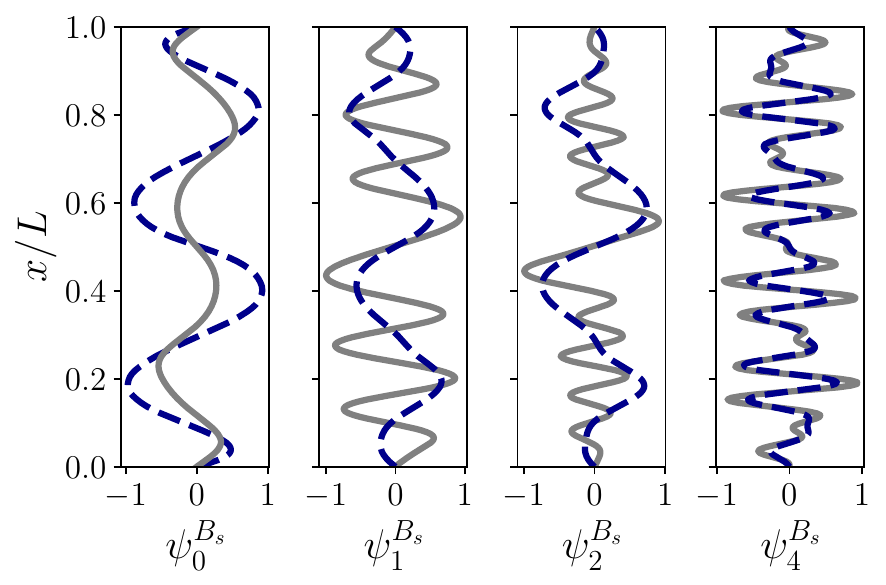}
         \caption{}
     \end{subfigure}
     \hfill

        \caption{ Mode identification of multiscale magnetic field $B_s$. (a) Amplitudes and frequencies of the DMD modes, compared with FFT of the data (dashed) of dynamo waves from~(\ref{eq:sys2_A}-\ref{eq:sys2_u}),  without imposing $\alpha$-effect (figure~\ref{fig:RiPr_kx13_dataset}a). Filled circles, standard DMD of rank $r=14$;  filled squares, Hankel DMD with $r=14$ and delay $d=20$;  empty squares, Hankel DMD, $r=4$, $d=20$. Dots corresponds to high-rank Hankel DMD, $r=40$, $d=300$.  Only positive DMD frequencies are shown. (b) FFT (dashed) and DMD spectrum of the model with imposed $\alpha$-effect (figure~\ref{fig:RiPr_kx13_dataset}b). Filled circles, standard DMD of $r=14$;  filled squares, Hankel DMD with rank $r=14$ and delay $d=9$;  empty squares, Hankel DMD, $r=6$, $d=9$; dots, high-rank Hankel DMD, $r=22$, $d=10$. Crossed-out circles correspond to spurious eigenvalues that appear in high-rank models ($r=22$) without increasing $d$. (c) Cumulative singular value spectrum of the multi-scale dynamo model with (circles) and without (solid lines) imposed  $\alpha$-effect.  Dark blue, $\sigma^2$; light blue, $\sigma$.  Vertical lines denote 99\% cut-off in both spectra ($\sum_i \sigma_i (\sigma_i^2)/\sum_{tot} \sigma (\sigma^2) = 0.99$ , dashed). (d) Profiles of DMD modes $\psi_0$, $\psi_1$, $\psi_2$ (slow cluster) and $\psi_4$ (fast cluster). Solid, with imposed $\alpha$-effect; dashed, without imposed $\alpha$-effect.}
        \label{fig:RiPr_kx13_modes}
\end{figure*}

\begin{table*}
    \centering
    \makebox[\textwidth][c]{
    \begin{tabular}{c|c|c|c|c|c|c|c|c|p{3cm}}
  &  &  \multicolumn{3}{|l|}{With imposed $\alpha$-effect} &  \multicolumn{4}{|l|}{Without imposed $\alpha$-effect} &  Interpretation \\ \hline \hline
  &  & DMD,  & HDMD & HDMD & FFT & DMD & HDMD & HDMD &  \\
   &  & $r=14$ & $r=6$, $d=9$ & $r=14$, $d=9$&  & $r=14$ & $r=14$, $d=20$ & $r=40$, $d=300$  &  \\
    \hline
 \parbox[t]{2mm}{\multirow{3}{*}{\vspace{-0.1cm}\rotatebox[origin=c]{90}{Slow}}}& $\omega_0$ &$ 0.282$ &$  0.282$ &$ 0.282$ & $0.209$ & $0.257$ &$0.244$& $0.229$& Dynamo wave, $\Im(\omega_0)$\\
& $\omega_1$& $0.845$ &  & $0.847$ &$0.670$ & & &$0.687$ &$\Im(\omega_1) \approx 3 \Im(\omega_0)$ \\
 & $\omega_2$&   $ 1.435$ &  & $1.426$ &$1.131$ &&$1.267$ & $1.145$& $\Im(\omega_2) \approx 5 \Im(\omega_0)$  \\
 & & & & &$1.591$ & $1.499$ & &$0.687$ &$\approx 7.6 \Im(\omega_0)$ \\
 & & & & &$2.052$ && & $2.061$& $\approx 9.9 \Im(\omega_0)$  \\
 & & & & &$2.512$&& & $2.519$& $\approx 12.0 \Im(\omega_0)$  \\
  \hline \hline 
 \parbox[t]{2mm}{\multirow{3}{*}{\vspace{-0.5cm}\rotatebox[origin=c]{90}{Fast}}}
 &  & &  &  & $11.850$ &$11.237$&$11.781$&$11.854$& $\Im(\omega_u) - \Im(\omega_2)$\\
& $\omega_3$ & $11.871$ &  & $12.152$ & $12.311$& & &$12.313$&$\Im(\omega_u) - \Im(\omega_1)$\\
& $\omega_4$ & $12.715$ & $12.715$ & $12.718$ & $12.772$ & $12.790$ &$12.758$&$12.771$&$\Im(\omega_u) - \Im(\omega_0)$\\
& $\omega_5$ & $13.279$ & $13.282$ & $13.282$ & $13.232$ & $13.263$&$13.238$& $13.229$&$\Im(\omega_u) + \Im(\omega_0)$\\
& $\omega_6$ & $13.694$ &  & $13.847$ & $13.693$ &&&$13.687$& $\Im(\omega_u) + \Im(\omega_1)$\\
&  & &  &  & $14.153$ &&$14.108$& $14.145$& $\Im(\omega_u) + \Im(\omega_2)$\\
\hline \hline 
& & && & &        &$26.031$&&Higher-order interaction\\
&   & & && & $5.404$, $3.567$ &      && Spurious\\
 \hline \hline
    \end{tabular}
    }
    \caption{Imaginary part of DMD eigenvalues  (frequencies) of  the total magnetic field $B_s = B +b $ from the dynamo models with and without imposed $\alpha$-effect, obtained with standard and Hankel DMD for different values of rank and delay. The parameters are specified in the header; the frequencies were rounded up to the third digit.  Only frequencies of peaks with the amplitude at least 0.001 of the dominant one are shown. FFT frequencies in the dynamo with imposed $\alpha$-effect are the same as those recovered with Hankel DMD with $r=14$, $d=9$. }
    \label{tab:DMD_eigs_RiPr_aug}
\end{table*}

As before, we perform dimension reduction of  the mixed magnetic field $B_s$ with POD. The largest singular values still come in pairs, as $B_s$ visibly retains some degree of symmetry. Circles in figure~\ref{fig:RiPr_kx13_modes}(c) denote the cumulative contribution of the first $i$ POD modes to the full data set, both in terms of $\sigma$ and $\sigma^2$. As expected, the cumulative energy of the modes $\sum_i \sigma_i^2$ grows faster than  $\sum_i \sigma_i$: only 5 modes are necessary to reproduce 99\% of energy in the flow, compared to 14 according to $\sigma$-criterion. However, with such a small $r$ only the principal dynamo wave with $\omega_0$ is captured correctly with DMD and spurious eigenvalues appear. We thus perform DMD on the POD basis of rank $r=14$. 

Two mode clusters are detected in the DMD spectrum of $B_s$:  low-frequency modes $\omega_{0-2}$ with $\Im (\omega)< 3 $ and high-frequency modes $\omega_{3-6}$ with  $\Im (\omega)> 10$ (figure~\ref{fig:RiPr_kx13_modes}b). The mode $\psi_0$ with the slowest oscillation frequency $\Im(\omega_0) \approx 0.282$ corresponds to the dynamo wave, with a shape similar to the Parker's dynamo wave (figure~\ref{fig:RiPr_modes}c). It is the large-scale oscillation visible in figure~\ref{fig:RiPr_kx13_dataset}. The other frequencies of the slow-evolution cluster,  $\Im(\omega_1) \approx \pm 3 \Im (\omega_0)$, $\Im(\omega_2) \approx \pm 5 \Im(\omega_0)$, correspond to the nonlinear interactions of the mode $\psi_0$ with itself through the nonlinearities $b^3$, $B^3$ in (\ref{eq:sys2_B}-\ref{eq:sys2_b}), as shown in section~\ref{sec:DMD_parkers}. The weaker frequency component  $\Im(\omega_2)$ appears through secondary interaction of $\omega_0$ and $\omega_1$ in this nonlinearity, consider, for example $[\sin (\omega_0) + \sin(3 \omega_0)]^3$. Despite being dynamically related, the characteristic spatial length scale of the modes $\psi_1$, $\psi_2$ is smaller then the length scale of the large-scale dynamo wave  $\psi_0$ (figure~\ref{fig:RiPr_kx13_modes}d). The eigenvalues in the second, ``fast", cluster are located around the forcing frequency $\omega_u = 13$, but none of them coincides with it exactly.  Instead, they are result of interaction of the modes in the slow-evolution cluster,  $\psi_{0,1}$, with the forcing $u$. Indeed, the nonlinear terms in model~(\ref{eq:sys2_A}-\ref{eq:sys2_b}) imply $b_t \propto  u_x B \propto \cos( \pm \omega_u t) \exp(\omega_{0,1} t) \propto \pm \omega_u \pm \Im(\omega_{0,1})$. 
The time-dependence of the DMD modes is exponential according to (\ref{eq:linsys_disc}) and (\ref{eq:DMD_modes}), so their interactions will contain all possible combinations of sine and cosine functions. Length scales of these faster oscillating modes are much smaller than length scales of the slow modes (see figure~\ref{fig:RiPr_kx13_modes}d).  

Comparison of DMD frequencies and amplitudes with  FFT results in figure~\ref{fig:RiPr_kx13_modes}(b) allows to evaluate  DMD performance. The optimal amplitudes of each DMD mode were calculated using the best-fit of the data matrix $Q$ onto the DMD modes \cite{jovanovic2014sparsity}. Although standard DMD of rank $r=14$ captures  both fast and slow frequency clusters, the agreement with FFT is imperfect. The DMD amplitudes of the modes $\omega_{2-6}$ are larger compared to the FFT amplitudes; weaker DMD eigenvalues of the ``fast" cluster, $\omega_3$ and $\omega_6$, are shifted towards slower frequencies; these less energetic modes are captured by standard DMD with less precision.

To improve DMD accuracy, we use the method of delays (Hankel DMD), as described in section \ref{sec:HankelDMD}. We increase the delay parameter up to $d=9$, and keep the rank $r=14$ to compare with the standard approximation. Hankel DMD identifies the same dynamical components of $B_s$ as standard DMD; however, the weak dynamical components with frequencies $\omega_2$, $\omega_3$ and $\omega_6$ are now in a better agreement with FFT results (figure~\ref{fig:RiPr_kx13_modes}b). The real parts of both weak and strong modes become considerably more neutral (see figure~\ref{fig:RiPr_aug_d_r_error}a).  Combined with reducing rank $r$, Hankel DMD leads to more compact  yet physical representation of the system dynamics. Consider a reduced-rank DMD decomposition, $r=6$, of the augmented Hankel matrix $Q_H$ with delay $d=9$. Although only one additional degree of freedom was allowed  compared to $r=5$ from $\sigma^2$ cut-off, the principal dynamical modes $\omega_0$, $\omega_4$, $\omega_5$ with the largest amplitudes are identified correctly (figure~\ref{fig:RiPr_kx13_modes}b). See table~\ref{tab:DMD_eigs_RiPr_aug}  for a list of detected DMD eigenvalues and  Appendix \ref{sec:accuracy_r_d} for further discussion on DMD accuracy as a function of rank and delay.

\subsection{DMD of the dynamo without imposed $\alpha$-effect}\label{sec:lin_validity}
After identifying the dominant interactions in the system with a well-defined $\alpha$-effect, it becomes straightforward to perform the same analysis on the original multiscale dynamo system~(\ref{eq:sys2_A}-\ref{eq:sys2_u}). We analyse the modified dynamo waves in figure~\ref{fig:RiPr_kx13_dataset}(a), discarding transients. As mentioned before, FFT of these data (figure~\ref{fig:RiPr_kx13_modes}a) also results in a ``slow" frequency cluster around the dynamo wave with $\Im(\omega'_0) = 0.209$ and the ``fast" one around $\omega_u$. Here primes are used to distinguish these frequencies from  those with imposed $\alpha$-effect in section~\ref{sec:RiPr_aug_DMD}. Analysis of the dominant frequency peaks of this spectrum in table~\ref{tab:DMD_eigs_RiPr_aug} shows that the ``fast" cluster corresponds to quadratic interactions of $\omega_u$ with magnetic field. Without \textit{a priori}  defined large-scale $\alpha$-effect, the spectrum of the ``slow" modes becomes more broad-band; nevertheless, principal interactions of the dynamo wave with itself corresponding to $\Im(3 \omega'_0)$,  $\Im(5\omega'_0)$, are still discernible. Higher frequencies of this cluster, $\Im(9.9 \omega'_0)$ , $\Im(12 \omega'_0)$ are potentially influenced by interactions through quadratic terms, but their magnitudes are weak.  After these interactions, the energy is further transferred downscale, and so additional weaker frequency peaks appear around $2 \omega_u = 26$ (table~\ref{tab:DMD_eigs_RiPr_aug}). As this frequency cluster has much lower energy, it was not shown in figure~\ref{fig:RiPr_kx13_modes}a.

The singular value spectra of this system are similar to those with imposed $\alpha$-effect (solid lines in figure~\ref{fig:RiPr_kx13_modes}c), with the difference that the first pair of POD modes contains more energy, as the amplitude of the large-scale field is stronger in this case (figure~\ref{fig:RiPr_kx13_dataset}a,c). It is thus sufficient to use $4$ POD modes according to $\sigma^2$- and $14$ modes according to $\sigma$-criterion for subsequent DMD decomposition. In figure~\ref{fig:RiPr_kx13_modes}(a), we compare resulting frequencies of DMD modes with the FFT spectrum of the field.  In this case, standard DMD decomposition with $r=14$ according to $\sigma$-criterion reliably identifies  only the large-scale dynamo wave with $\Im(\omega_0)$ and two strongest modal pairs of its quadratic interaction with the flow, $\Im(\omega_4)$, $\Im(\omega_5)$. The rest of the DMD eigenvalues either approximate several dynamical components of the spectrum when located in between dominant peaks, or do not correspond to energetic flow structures and thus are spurious (table~\ref{tab:DMD_eigs_RiPr_aug}). In this case, using Hankel DMD is crucial to improve scale detection. Using delay of $d=20$ allows to resolve the ``slow" frequency $5 \Im (\omega_0)$ and its interactions with the flow from the ``slow" cluster, as well as to approximate second-order quadratic interactions proportional to $2 \omega_u$. As previously, Hankel DMD with $d=20$ allows to use low, $\sigma^2$ model rank $r=4$ and  to capture both fast and slow dynamics using only two pairs of complex-conjugate modes. Finally, increasing further the values of rank and delay, to $r=40$ and $d=300$, allows to detect all energetic spectral peaks in figure~\ref{fig:RiPr_kx13_modes}(a).

Spatial shapes of the modes in the ``fast" and the ``slow" clusters highlight some differences in dynamics of the two systems. In the system without imposed $\alpha$-effect, DMD modes are slightly more symmetric with respect to $x/L=0.5$, as $b$  has much less pronounced systematic time-periodic component (figure~\ref{fig:RiPr_kx13_dataset}). The first two complex-conjugate DMD modes with the dominant frequency $\Im(\omega_0)$ are able to reproduce the spatial modulation in figure~\ref{fig:RiPr_kx13_dataset}(a), indicating that it is related to the spatial shape of filtered $\alpha$-effect. As the nonlinearity $ \gamma b^3$ was set to zero in this model, the ``slow" modes $\omega_0$, $\omega_1$ and $\omega_2$ remain large-scale (figure~\ref{fig:RiPr_kx13_modes}d), while the spatial distribution of small-scale modes from the ``fast" cluster is nearly identical both dynamo systems. Except for these features, the systems with and without imposed $\alpha$-effect have qualitatively similar dynamics.

\section{Validity of linear approximation}\label{sec:lin_validity}

\begin{figure*}
     \centering
     \begin{subfigure}[b]{0.48\textwidth}
         \centering
         \includegraphics[width=\textwidth]{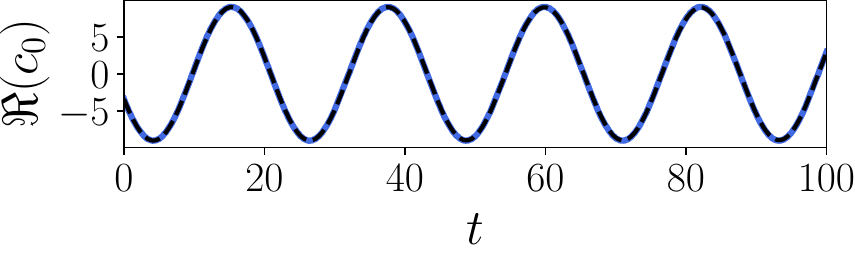}
         \caption{}
     \end{subfigure}
     \hfill
     \begin{subfigure}[b]{0.48\textwidth}
         \centering
         \includegraphics[width=\textwidth]{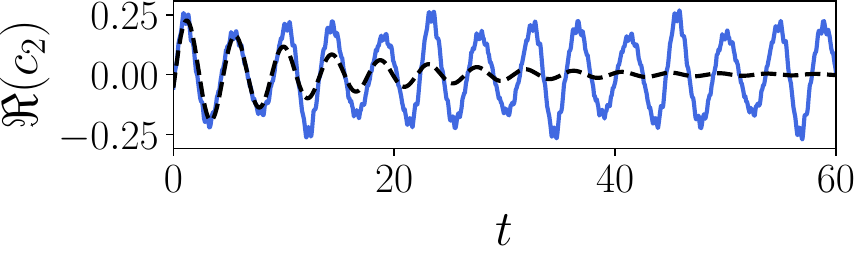}
         \caption{}
     \end{subfigure}
     \vfill 
          \begin{subfigure}[b]{0.48\textwidth}
         \centering
         \includegraphics[width=\textwidth]{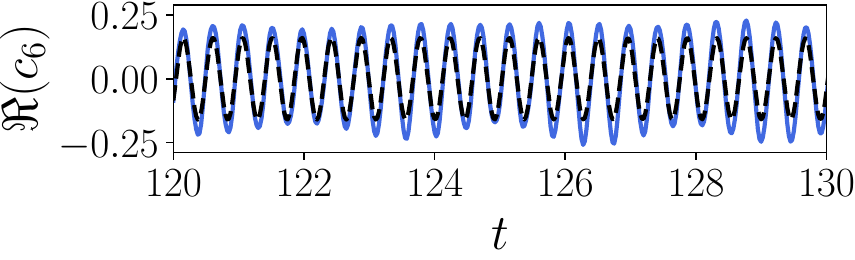}
         \caption{}
     \end{subfigure}
     \hfill
     \begin{subfigure}[b]{0.48\textwidth}
         \centering
         \includegraphics[width=\textwidth]{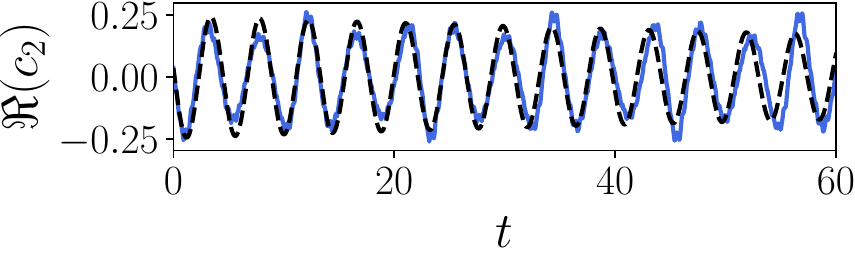}
         \caption{}
     \end{subfigure}
     \vfill
    \begin{subfigure}[b]{\textwidth}
         \centering
         \includegraphics[width=\textwidth]{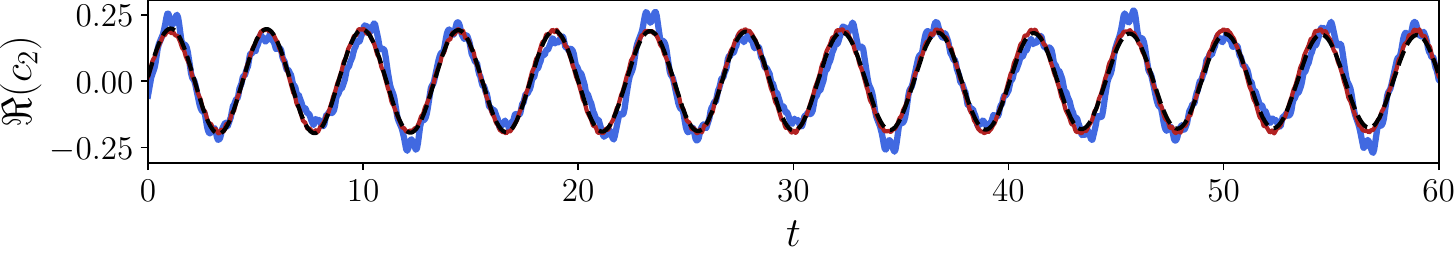}
         \caption{}
     \end{subfigure}
        \caption{Temporal evolution of DMD modes as a function of time. Solid, instantaneous (actual) coefficients $c_i^I$; dashed, coefficients of the linear model $c_i^L$. Only real part of $c_i$ is shown.  (a) $c_0$, dynamo wave, standard DMD with $r=14$. (b) $c_2$, corresponding to the mode  $\psi_2$, standard DMD with $r=14$. (c) $c_6$, corresponding to the mode $\psi_6$, Hankel DMD, $r=14$, $d=9$. (d) $c_2$, Hankel DMD, $r=14$, $d=9$.  (e) $c_2$, Hankel DMD,  $r=22$, $d=10$ (dashed, in red); $c_2^I$ obtained with standard DMD  is given in blue for comparison. See table~\ref{tab:DMD_eigs_RiPr_aug} and figures~\ref{fig:RiPr_kx13_modes}(b-d) for mode description. }
        \label{fig:RiPr_aug_d_r_coef}
\end{figure*}

In the following, we analyse how Hankel DMD improves scale detection on the data from multi-scale dynamo system with imposed $\alpha$-effect~(\ref{eq:sys2_A}-\ref{eq:sys2_emf}), (\ref{eq:RiPr_u}-\ref{eq:RiPr_aug_emf}). As this system has fewer dynamical components (figure~\ref{fig:RiPr_kx13_modes}b), it allows to track easily the influence of varying parameters of DMD model on detection and properties of individual  modes without increasing computational complexity. We leave the corresponding analysis of the multi-scale dynamo without imposed $\alpha$-effect for the future work.

DMD approximates the flow dynamics with a linear system; however, the actual temporal evolution of  DMD modes can be more complex than the one prescribed by~(\ref{eq:linsys_disc}). According to~(\ref{eq:linsys_disc}) and \eqref{eq:DMD_modes}, temporal coefficients $c_i^L(t)$ of the DMD modes  $\psi_i$ are defined by DMD eigenvalues through $c_i^L(t) =\exp(\omega_i t) b_i(0)$. This linear model allows either growing or decaying 
 in time oscillating modes. However, we can also evaluate actual instantaneous temporal coefficients of the modes, $ c_i^I(t)$, from the data. This is done using non-orthogonal projection of the data matrix $Q$ on the matrix of DMD modes $\Psi$, 
\begin{equation}\label{eq:inst_cI}
    \bm{c}^I(t) = \Psi^\dag Q.
\end{equation}
where $\Psi^\dag$ denotes its pseudoinverse.  By comparing  $ c_i^I(t)$ and $c_i^L(t)$, we can assess how linear is the dynamics of the $i$-th DMD mode.

Figure~\ref{fig:RiPr_aug_d_r_coef}(a) compares $ c_0^I(t)$ and $c_0^L(t)$  of the dynamo wave mode $\psi_0$. Excellent agreement between the two indicates that the dynamo wave component evolves essentially linearly. The energetic modes from the ``fast"  cluster, $\psi_4$ and $\psi_5$, also feature nearly linear dynamics but faster frequencies (not shown here). These modes are represented relatively accurately both by DMD and Hankel DMD. On the other hand, the actual temporal dynamics of  weaker modes $\psi_2$, $\psi_3$ and $\psi_6$ in both ``slow" and ``fast" clusters is not entirely linear. In figure~\ref{fig:RiPr_aug_d_r_coef}(b) we illustrate this on the temporal coefficient $c_2^I(t)$ of the mode with $\omega_2$. This mode, besides the principal oscillating frequency, exhibits a slower modulation on the time scale comparable to the one of $\psi_0$. When $|c_2^I (t)|$ is large, it is contaminated by a faster evolving signal. The corresponding linear coefficient $c_2^L(t)$ is identified by the standard DMD as decaying (see negative real parts of the corresponding DMD eigenvalues in figure~\ref{fig:RiPr_aug_d_r_error}a), and rapidly becomes out-of-phase with the actual coefficient $c_2^I(t)$. The $\psi_2$ component is thus quickly lost in the linear DMD model.

\citet{callaham2022role} showed that POD and DMD modes of a flow with two principal frequencies are nonlinearly correlated through  triadic interactions, creating mixed-frequency content. In our augmented system, the nonlinearity is more complex, with both quadratic and cubic terms. Quantitative estimation of these correlations is thus out of scope of this work. Nevertheless, we identify the presence of mixed frequency dynamics from the temporal coefficients qualitatively. From equations~(\ref{eq:sys2_B}-\ref{eq:sys2_b}), the mode with eigenvalue $\omega_2$ appears through cubic interaction between the dynamo wave $\omega_0$ and its first nonlinearity with $\Im(\omega_1) = 3\Im(\omega_0)$ as $[\exp (\omega_0) + \exp(3 \omega_0)]^3$. When expanded, this gives the following set of harmonics: $ [\exp (3 \omega_0) + 3 \exp (5 \omega_0) + 3 \exp (7 \omega_0) + \exp (9 \omega_0)]$. In the FFT of $c_2^I(t)$ (not shown for brevity), all these harmonics are present, and the frequency content of $7 \Im(\omega_0)$ is only 10 times weaker than the principal signal frequency, $\Im(\omega_2) \approx 5 \Im(\omega_0)$. The fast-oscillating contribution to the time series of $c_2^I(t)$ corresponds to the frequencies of  $\Im(\omega_u \pm \omega_2)$, i.e., interaction of $\psi_2$ and the small-scale part of the flow $u$. Thus, DMD decomposition of rank $r=14$ mixes these dynamical components in a single one.  The weak modes $\psi_3$, $\psi_6$ from the ``fast" modal cluster are also modulated, although less then $\psi_2$ (figure~\ref{fig:RiPr_aug_d_r_coef}c). 

We introduced delay in the data with Hankel DMD to see whether this  prediction improves, while keeping the model rank $r=14$. It can influence the results in two ways: first, by identifying modal frequency and amplitude more precisely; and second, by obtaining a spatial modal basis that is closer to Koopman eigenvectors of the system. In our case, the changes in spatial shape of mode $\psi_2(x)$ were minor; but its eigenvalue $\omega_2$  was identified as less dampened (figure~\ref{fig:RiPr_aug_d_r_error}a). The mode now decays much slower and its linear temporal evolution represents the real system well at early times (figure~\ref{fig:RiPr_aug_d_r_coef}d). Eventually, it gets out of phase with the instantaneous modal coefficient, indicating that the temporal signal of $\psi_2$-mode is still mixed-frequency. 
Hankel DMD also recovers modes with $\omega_3$ and $\omega_6$ as nearly non-decaying and in-phase with the actual signal (figure~\ref{fig:RiPr_aug_d_r_coef}c), when standard DMD would result in rapidly decaying, out-of-phase behaviour similar to figure~\ref{fig:RiPr_aug_d_r_coef}(b).

Further improvement is achieved by increasing rank $r$ beyond the one given by $\sigma$-criterion, as shown by high-rank DMD results in figure~\ref{fig:RiPr_kx13_modes}b (red dots). The rank of $r=20$ and delay $d=6$ are large enough to separate the weakest dynamical components with $7\Im(\omega_0)$,  $\Im(\omega_u \pm \omega_2)$ from the mode $\psi_2$. Then, the temporal coefficient $c_2^I(t)$ becomes uncorrelated with them (figure~\ref{fig:RiPr_aug_d_r_coef}e), the corresponding mode $\psi_2$ behaves linearly and according to (\ref{eq:linsys_disc}). With smaller delays, these additional dynamical components are not recovered and spurious eigenvalues appear (crossed-out circles in figure~\ref{fig:RiPr_kx13_modes}b).  


\section{Conclusions}\label{sec:discussion}
In this work, we used Proper Orthogonal Decomposition and Dynamic Mode Decomposition for scale identification in one-dimensional dynamo models. Our first benchmark model belongs to a family of $\alpha-\Omega$ dynamos  suggested by \citet{parker1955hydromagnetic}, \citet{proctor2007effects} and \citet{richardson2010effects}, among others. We found that POD and DMD modes of this system have similar spatial shapes; in fact, DMD can be viewed as a rotation of the POD basis \citep{callaham2022role}. However, complex-conjugate DMD modes give a sparser dynamical basis for the oscillating dynamo. We identified the modes corresponding to the dynamo wave, and  the modes appearing due to the nonlinear damping term $B^3$ in the model.  Together, these two DMD modes form an accurate linear model of the dynamo. 

The shortcoming of the benchmark dynamo model is that it has only one independent dynamo wave component. We constructed two \textit{qualitative} augmented dynamo models featuring both the dynamo wave and the small-scale magnetic field forced by the flow, with and without imposed $\alpha$-effect. The aim of these reduced models was not to reproduce the mean-field dynamo theory in a predictive way but rather to create an clear benchmark for DMD analysis in multiscale dynamos. By choosing the frequency of the flow, we separated dynamically relevant scales in space and time in both models.  DMD eigenvalues of this dynamo were located in two regions of the complex plane:  the ``slow" cluster near the dynamo wave with frequency $\Im(\omega_0)$, and the ``fast" cluster, centered about the small-scale flow frequency $\omega_u$. Analysis of the frequency distribution suggests that the ``slow" cluster results from the nonlinear interactions of the dynamo wave with itself through the cubic terms $B^3$, $b^3$ in equations (\ref{eq:sys2_B}-\ref{eq:sys2_b}), while the ``fast" cluster appears through quadratic nonlinearity proportional to the product of $u$ and $B$ (or $b$) in equations \eqref{eq:sys2_A}, \eqref{eq:sys2_b}. As nonlinearities in the models with and without imposed $\alpha$-effect have similar functional form, the temporal dynamics of both models is qualitatively similar. Note that DMD is not able to separate $B$ and $b$ from the mixed signal $B_s$; instead, it identifies different scales with the same temporal dynamics as a single mode. Thus, DMD modes from the ``slow" cluster in figure~\ref{fig:RiPr_kx13_modes}(d) are slightly asymmetric and take into account the systematic large-scale footprint of the dynamo wave in $b$ through  term $u_x B$. In the dynamo without imposed $\alpha$-effect, the modes corresponding to cubic nonlinear interactions remain uncontaminated with  small scales, because cubic nonlinearity for $b$ was suppressed in this model, $\gamma=0$ compared to $\gamma=10$ in the case with imposed $\alpha$-effect. The model without imposed $\alpha$-effect thus features a better separation of spatial scales. 

Furthermore, we investigated the influence of the DMD parameters, rank and delay, on accuracy of scale detection.
Fast-scale dynamics is not captured if a more conservative $\sigma^2$-criterion for $r$ is applied. Thus, rank $r$ is important if  dynamically relevant modes are not  the most energetic ones. We demonstrated that Hankel (high-order) DMD improves the accuracy of approximation, recovering the amplitudes and frequencies of weak dynamical components in the data with more precision. In the dynamo with imposed $\alpha$-effect, we identified that the linearity assumption breaks for the weakest DMD modes; Hankel DMD gives a better linear representation for these modes, especially if the rank of the model is large enough to separate mixed-frequency components in the these modes. This improvement relies on spatiotemporal coherency of the corresponding dynamical scales. In dynamical systems with a more broad-band spectrum,  like the one without imposed $\alpha$-effect, the errors of linear model~\eqref{eq:linsys_disc} will be larger; nevertheless, Hankel DMD still considerably improves accuracy of detection and modelling of energetic length scales. In this case, further extensions of DMD method are of interest - for example, the multiresolution DMD based on wavelet-like techniques proposed by \citet{dylewsky2019dynamic}. We leave this analysis for the future work.

Finally, we give some notes on extension of this approach to more realistic DNS of shear dynamos. There, the small-scale  flow  is frequently forced by a helical forcing with a defined length scale, or forcing wave number, but randomized in time \citep{tobias2013shear,nigro2017large}. In this case, DMD is expected to represent poorly the small-scale flow and field components, because they are incoherent in time. Nevertheless, it should be possible to extract the large-scale, coherent components of the magnetic and velocity fields corresponding to the ``slow" modal cluster from such simulations. As a final remark, our augmented benchmark dynamo model has a well-defined scale separation while many turbulent dynamos have continuous energy spectra. These dynamos are expected to give continuous DMD spectra rather than clustered. 

\section*{Acknowledgements}
This project has received funding from the European Union’s Horizon 2020 research and innovation programme under the Marie Skłodowska-Curie grant agreement No 890847. The author is grateful to Prof. Steven Tobias and Dr. Calum Skene for discussions that helped to shape many aspects of this work. Support from the ARC supercomputing center at the University of Leeds, Leeds Institute for Fluid Dynamics, and the Isaac Newton Institute for Mathematical Sciences, Cambridge, during the programme ``Frontiers in dynamo theory: from the Earth to the stars", is fully acknowledged.  This work was supported by EPSRC grant no EP/R014604/1.

\section*{Data Availability}
The data in this work were generated using publicly available code Dedalus~\citep{burns2020dedalus} and will be provided on request to the corresponding author. 
\bibliographystyle{mnras}
\bibliography{dynbib}

\appendix

\section{Influence of DMD rank and delay }\label{sec:accuracy_r_d}

\begin{figure*}
     \centering
     \begin{subfigure}[b]{0.45\textwidth}
         \centering
         \includegraphics[width=\textwidth]{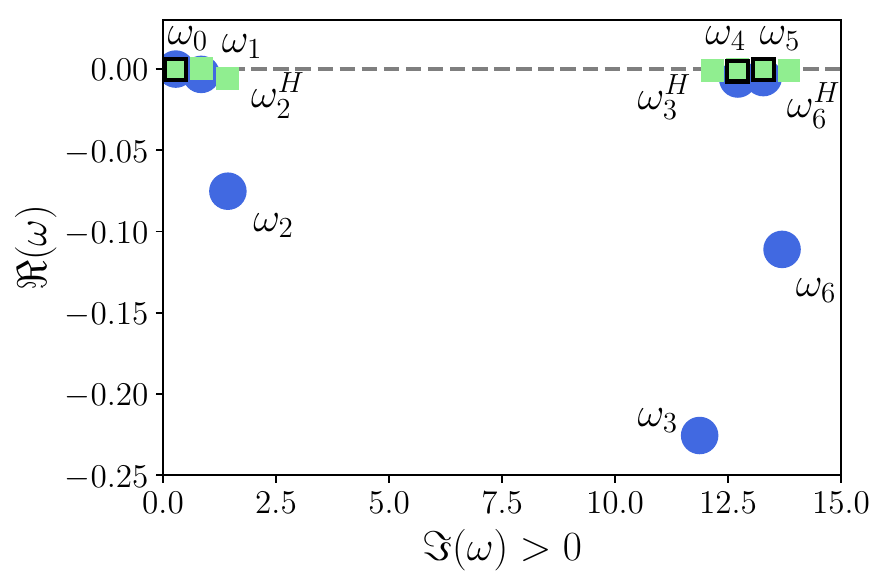}
         \caption{}
     \end{subfigure}
     \hfill 
          \begin{subfigure}[b]{0.45\textwidth}
         \centering
         \includegraphics[width=\textwidth]{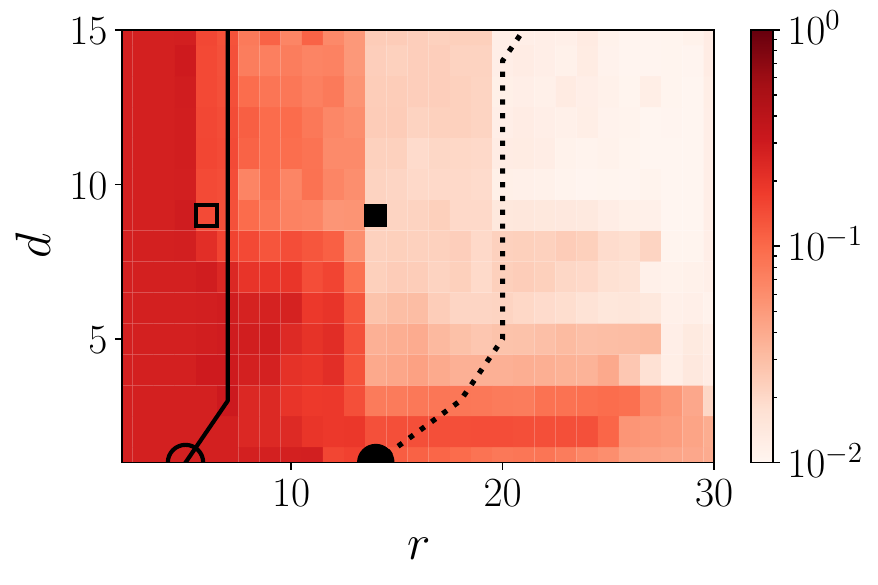}
         \caption{}
     \end{subfigure}
        \caption{ (a) Spectrum of DMD eigenvalues of the multiscale dynamo model without imposed $\alpha$-effect; only eigenvalues with positive imaginary part (frequency) are shown. Filled circles, standard DMD of $r=14$;  filled squares, Hankel DMD with rank $r=14$ and delay $d=9$;  empty squares, Hankel DMD, $r=6$, $d=9$. Dashed, neutral stability line. Superscipt $H$ denotes frequencies obtained with Hankel DMD.  (b) The  model error  $||B_s - B_s^{model}||_2/||B_s||_2$ in this model as a function of rank and delay. The colours are spaced logarithmically.  Symbols represent the  analysis parameters in panel (a). The empty circle corresponds to standard DMD with $r=5$ according to $\sigma^2$-criterion. Solid line, $\sigma^2$-criterion;  dotted, $\sigma$-criterion. } 
        \label{fig:RiPr_aug_d_r_error}
\end{figure*}

In this section, we show how varying the rank and delay influences accuracy of the linear DMD model, on the example of augmented dynamo with imposed $\alpha$-effect (\ref{eq:sys2_A}-\ref{eq:sys2_b}), (\ref{eq:RiPr_u}-\ref{eq:RiPr_aug_emf}). As the dynamics in this system is predominantly periodic, one would expect the DMD modes computed in the data set to be nearly neutral (i.e. with $\Re(\omega) \approx 0$). However, DMD eigenvalues $\omega_2$, $\omega_3$ and $\omega_6$, computed with the standard DMD algorithm with rank $r=14$ according to $\sigma$-criterion, are dampened and have pronounced negative growth rates (figure~\ref{fig:RiPr_aug_d_r_error}a). This will contribute to overall error~\eqref{eq:lin_error} of the linear DMD approximation~\eqref{eq:linsys_disc}, since heavily dampened components will decay rapidly. On the other hand, Hankel DMD with delay $d=9$ results in all the modes being close to neutral stability, reducing the error from 12\% to 2\%.

Overall, there is no analytical expression  for optimal $d$ and $r$ which tend to increase with the data complexity. So far,  we performed a point-by-point analysis in the parameter space of $(r,d)$; it can be generalized by spanning a range of $r$ and $d$ and tracking error~(\ref{eq:lin_error}) of linear DMD approximation~(\ref{eq:linsys_disc}). Figure~\ref{fig:RiPr_aug_d_r_error}(b) presents this error as a function of both $r$ and $d$, with the DMD models corresponding to figures~\ref{fig:RiPr_aug_d_r_error}(a) denoted accordingly.  With or without delay, increasing rank reduces the error of the linear approximation. Increasing delay improves accuracy at first, but when $d$ is large, the dependence of error on $d$ saturates. To get a better understanding of this behaviour, we evaluated how delay affects the rank of Hankel data matrix $Q_H$~(\ref{eq:HankelQ}). We evaluate this in terms of  rank $r^{\sigma (\sigma^2)}$  of $Q_H$ according to either $\sigma$ or $\sigma^2$ criteria. These ranks are denoted in figure~\ref{fig:RiPr_aug_d_r_error}(b) by dotted and solid lines, respectively. They estimate the number of independent vector components required to describe certain percentage of the information contained in $Q_H$. At first,  their values increase with $d$, indicating increase in data rank as newly added delayed system states introduce new dynamical information. However, this trend saturates around $d=5$. Further increasing delay does not introduce new dynamical components, and the representation error becomes predominantly a function of $r$ when $d$ is large. A reasonable minimum bound for $d$ therefore is that the rank of matrix $Q_H$, defined either through $\sigma$, $\sigma^2$ criteria or numerical estimation of the matrix rank, should saturate. 

Another insight from figure~\ref{fig:RiPr_aug_d_r_error}(b) is that increasing delay does not recover the dynamical components that were entirely removed by truncating the POD modal basis at rank $r$, as demonstrated by the abrupt decrease in error between $r=13$ and $r=14$, from $\epsilon \sim 0.1$ to $\epsilon \sim 0.01$. At $r=14$, the weak DMD harmonic with frequency $\omega_2$ first appears, improving accuracy. When $r<14$, this dynamical component is not identified by  DMD, independently of $d$. Note that FFT of $B_s$ shows other, even weaker frequencies of $\Im(7\omega_0)$,  $\Im(\omega_u \pm \omega_2)$ in figure~\ref{fig:RiPr_kx13_modes}(b), not captured by standard or delayed DMD with $r=14$. As a final note, using increasingly high delays like those used in figure~\ref{fig:RiPr_kx13_modes}(a) can result in eigenvalues moving up in the complex plane of figure~\ref{fig:RiPr_aug_d_r_error}(a) so that $\Re(\omega)>0$ and therefore some DMD modes become exponentially growing. In this case, increased accuracy of frequency detection will be accompanied by higher errors in the linear model~\eqref{eq:linsys_disc}.
\end{document}